\documentclass[12pt,a4paper]{article}
\usepackage{jheppub}
\usepackage{amsmath}
\usepackage{amsfonts}
\usepackage{amssymb}
\usepackage{graphicx}
\usepackage{mathrsfs}
\usepackage{bbold}

\newcommand{\state}[1]{\left|#1\right>}
\newcommand{\phy}{\state{\textrm{phys}}}

\newcommand{\bstate}[1]{\left<\bar{#1}\right|}
\newcommand{\bvac}{\bstate{\textrm{0}}}

\newcommand{\tr}{\textrm{tr}}
\newcommand{\Det}{\textrm{Det}}
\newcommand{\nCr}[2]{\begin{pmatrix}#1 \\ #2\end{pmatrix}}

\usepackage[]{youngtab}

\makeatletter
\DeclareRobustCommand{\rvdots}{%
  \vbox{
    \baselineskip4\p@\lineskiplimit\z@
    \vskip0.1em
    \hbox{.}\hbox{.}
  }}
  \DeclareRobustCommand{\rhdots}{%
  \mbox{   
    \hskip0.1em
   $\cdot\cdot$
  }}
\makeatother

\def\Tiny{\fontsize{5pt}{5pt} \selectfont}
\def\Small{\fontsize{8pt}{8pt} \selectfont}

\title{Mixed symmetry tensors in the worldline formalism}

\author[a,b]{Olindo Corradini,}
\author[c]{James P. Edwards}

\affiliation[a]{Dipartimento di Scienze Fisiche, Informatiche e Matematiche,\\ 
Universit\`a degli Studi di Modena e Reggio Emilia,\\ via Campi 213/A, I-41125 Modena, Italy} 
\affiliation[b]{INFN, Sezione di Bologna,\\ via Irnerio 46, I-40126 Bologna, Italy}
\emailAdd{olindo.corradini@unimore.it}

\affiliation[c]{Department of Mathematical Sciences, \\ 
University of Bath\\
Claverton Down, Bath\\
BA2 7AY, UK}
\emailAdd{jpe28@bath.ac.uk}

\abstract{We consider the first quantised approach to quantum field theory coupled to a non-Abelian gauge field. Representing the colour degrees of freedom with a single family of auxiliary variables the matter field transforms in a reducible representation of the gauge group which -- by adding a suitable Chern-Simons term to the particle action -- can be projected onto a chosen fully (anti-)symmetric representation. By considering $F$ families of auxiliary variables, we describe how to extend the model to arbitrary tensor products of $F$ reducible representations, which realises a $U(F)$ ``flavour'' symmetry on the worldline particle model. Gauging this symmetry allows the introduction of constraints on the Hilbert space of the colour fields which can be used to project onto an arbitrary irreducible representation, specified by a certain Young Tableau. In particular the occupation numbers of the wavefunction -- i.e. the lengths of the columns (rows) of the Young Tableau -- are fixed through the introduction of Chern-Simons terms. We verify this projection by calculating the number of colour degrees of freedom associated to the matter field. We suggest that, using the worldline approach to quantum field theory, this mechanism will allow the calculation of one-loop scattering amplitudes with the virtual particle in an arbitrary representation of the gauge group.}

\keywords{Gauge Symmetry, Scattering Amplitudes, Wilson loops}
\begin{document}
\maketitle
\section{Introduction}
The worldline formalism \cite{Strass1, Schu} is proving to be a valuable alternative to the more traditional approaches to quantum field theory. In particular it provides computationally efficient methods of obtaining scattering amplitudes, having been originally applied to re-derive the Bern-Kosower master formulae \cite{Bern, Sato} without recourse to string theory. It has now been applied to many problems, such as multi-loop amplitudes \cite{Dunne}, the analysis of effective actions \cite{Schmidt}, field theory on curved space time~\cite{Bastianelli:2002fv} and photon-graviton mixing \cite{Bgrav1}, higher spin fields~\cite{hspin2, hspin3} and tree-level scattering \cite{Me2un, Ahmadiniaz:2015xoa}, amongst many others. There are also phenomenological applications such as in photon / gluon scattering~\cite{gluon}, bound states computations~\cite{Bastianelli:2014bfa}, and the description of standard model matter \cite{Paul2} and its unification \cite{Me1}. These first quantised theories involve actions which realise a worldline supersymmetry, with the super-partners to the worldline coordinates providing the field's spin degrees of freedom. The interaction between the matter field and the gauge field is expressed through the gauge invariant Wilson-loop coupling.


More recently the worldline approach to non-Abelian field theory has been improved~\cite{Bastwl1, Bastwl2}, and the difficulty related to the path ordering of the Wilson loop (necessary to guarantee the gauge invariance) has been lifted, thus making the perturbative expansion more manageable. 
This was achieved by representing colour degrees of freedom (and the coupling to the Lie algebra valued gauge field) with additional auxiliary fields \cite{grass, grass2, Hoker}. These fields enlarge the Hilbert space of the quantum theory but lead to the matter field transforming in a reducible representation of the gauge group. The Hilbert space may be finite-dimensional or infinite dimensional, depending on how the colour degrees of freedom have been included in the worldline theory. Either way, projection onto a finite-dimensional  irreducible representation has been achieved by coupling the colour variables to a $U(1)$ gauge field, including an additional Chern-Simons term whose level is fixed so as to select the required representation \cite{Bastvt, Bastforms2}. The worldline gauge field does not modify the theory in any other way than to impose the constraint enforcing irreducibility, since in one dimension the $U(1)$ theory has trivial dynamics. By tuning the Chern-Simons coupling constant it is possible to pick out each representation at will.

One of the limitations that previous approaches have encountered is that it has only been possible to describe fully anti-symmetric or fully symmetric representations of the gauge group. This has followed from the use of additional fields which are anti-commuting (Grassmann variables) or commuting respectively. This is certainly sufficient to describe standard model fermions -- which transform in (conjugate-)fundamental or trivial representations~\cite{Paul2} -- and the multiplets used in $SU(5)$ unification -- which are fully anti-symmetric~\cite{Me1} -- but more exotic matter and a worldline description of gluons require a generalisation of these ideas. It is also interesting to uncover the additional structure the worldline theory requires in order to accommodate the description of more complex matter. In this article we intend to explain how this generalisation can be achieved by introducing several ``families'' of the auxiliary fields and partially gauging a unitary symmetry that rotates between them.

The Hilbert space that arises will be shown to be the tensor product of the spaces associated to each set of fields which provides a much richer structure. For this reason a more general procedure of projecting onto an irreducible representation is required which we will construct. In the present article we first deal with the case that the colour fields are fermionic, before repeating our analysis in the event that they are bosonic so as to provide a complete picture of our approach. For both scenarios we will study the phase-space theory that is familiar from the worldline approach and describe the introduction of the colour fields and the projection onto an arbitrary irreducible representation of the gauge group. To demonstrate that our construction is correct we will functionally quantise the resulting theories on the circle, using the path integral to compute the number of degrees of freedom associated to the matter field.

This article begins by revising the quantum mechanical description of spinning point particles and the incorporation of a coupling to the gauge field. In Section~\ref{SecExtension} we then discuss additional anti-commuting worldline fields that carry the particle's colour information, and the extra worldline global symmetry that arises. We achieve irreducibility by partially gauging this symmetry. In Section~\ref{SecPath} we carry out the functional quantisation of the colour fields to count the number of degrees of freedom they associate to the matter field. Having achieved this for fermionic colour fields, in Sections~\ref{SecTheoryB} and~\ref{SecGeneral} we briefly provide an account of the worldline theory for bosonic auxiliary fields, again quantising the theory on the circle to count the number of degrees of freedom in Section~\ref{SecPathB}.

\section{Worldline theory and anti-commuting colour fields}
\label{SecTheory}
First quantised approaches to quantum field theory begin with a quantum mechanical description of the dynamical and spin degrees of freedom of the field. For example, the phase-space description of a classical spin 1/2 point particle in Minkowski space 
(which is relevant for a worldline description of the Dirac field) takes the following form \cite{BdVH, BDZ}
\begin{equation}
	S\left[\omega, p, \psi, e, \chi \right] = \int_{0}^{1} d\tau \,\bigg[ p \cdot \dot{\omega} + \frac{i}{2}\psi \cdot \dot{\psi} -eH - i \chi Q\bigg],
	\label{N=1}
\end{equation}
where the $\omega^{\mu}$ represent the embedding of the particle's worldline in space and the $\psi^{\mu}$ describe its spin degrees of freedom. We have also introduced the functions $H = \frac{1}{2} p^{2}$ (Hamiltonian) and $Q = p \cdot \psi$ (super-charge) which are constrained to vanish by the equations of motion of the einbein, $e(\tau)$, and the gravitino, $\chi(\tau)$, respectively. Through Poisson (Dirac) brackets\footnote{We take $\{x^{\mu}, p^{\nu}\}_{PB} = \delta^{\mu\nu}$ and $\{\psi^{\mu}, \psi^{\nu}\}_{PB} = -i\delta^{\mu\nu}$ which generate gauge transformations through $\delta z = \big\{ z, \xi(\tau) H+ i\eta(\tau) Q \big\}_{PB}$.	} these functions generate the local super-symmetry transformations of the matter fields. 
\begin{equation}
	\delta \omega^{\mu} = \xi(\tau) p^{\mu} + i\eta(\tau)\psi^\mu(\tau); \qquad \delta p^{\mu} = 0; \qquad \delta \psi^{\mu} = -\eta(\tau)p^{\mu}
\end{equation}
where $\xi(\tau)$ is the generator of reparameterisations and $\eta(\tau)$ is the (Grassmann) generator of local supersymmetry transformations. Invariance of the action is achieved through the accompanying transformations of the super-gravity multiplet:
\begin{equation}
	\delta e = \dot{\xi}(\tau) + 2 i \chi \eta(\tau); \qquad \delta \chi  = \dot{\eta}(\tau),
\end{equation}
whilst the constraint functions satisfy the $\mathcal{N}  = 1$ supersymmetry algebra
\begin{equation}
	\left\lbrace Q, Q \right\rbrace_{_{PB}} = -2i H.
	\label{Qsusy}
\end{equation}
In the canonical quantisation scheme, the phase space variables are promoted to linear operators with the fundamental relations $\left[\hat{x}^{\mu}, \hat{p}^{\mu}\right] = i\delta^{\mu\nu}$ and $\{ \hat{\psi}^{\mu}, \hat{\psi}^{\nu} \} = \delta^{\mu\nu}$ and the constraints must be imposed as operator equations which define the physical state space: $\hat{p}^{2} \phy = 0$ and $\hat{\psi} \cdot \hat{p} \phy = 0$. The anti-commutation algebra can be solved by taking $\psi^{\mu} = \frac{1}{\sqrt{2}}\gamma^{\mu}$, which shows that the purpose of the spin degrees of freedom is to generate the $\gamma$-matrices. The constraints are then seen to ensure that the physical states are on shell and satisfy the Dirac equation.

To couple this particle (and consequently the underlying field which it is representing) to an Abelian gauge potential requires the modification of $H$ and $Q$. Minimal coupling is achieved by replacing the conjugate momentum, $p^{\mu}$, by its covariant version, $\pi^{\mu} = p^{\mu} - A^{\mu}$, where we have absorbed the coupling strength into the gauge field $A^{\mu}$. The effect of this is to the re-define the susy generator as $Q \equiv \psi \cdot \pi = \psi \cdot (p - A)$. Now the Hamiltonian, $H$, is determined by the supersymmetry algebra (\ref{Qsusy}) where a further change is encountered:
\begin{equation}
	\left\lbrace Q, Q \right\rbrace_{_{PB}} = -2i H \Rightarrow H = \frac{1}{2}\pi^{2} + \frac{i}{2}\psi^{\mu}F_{\mu\nu}\psi^{\nu},
\end{equation}
where $F_{\mu\nu}$ is the field strength tensor built out of $A_{\mu}$ as $F_{\mu\nu} = \partial_{\mu}A_{\nu} - \partial_{\nu}A_{\mu}$. This idea can be extended to the non-Abelian case, where the vector potential is Lie algebra valued. We will take the generators of this algebra, $\{T^{a}\}$, to be Hermitian and choose them in the fundamental representation of the symmetry group 
(we limit attention to the gauge group $SU(N)$ for physical reasons), so $A^{\mu} = A^{a\mu} T^{a}$. These additional details can be incorporated into the worldline action at a classical level by introducing additional Grassmann variables which carry the colour degrees of freedom that upon quantisation will create the associated Hilbert space~\cite{grass, grass2}. To do this we follow \cite{Bastwl1} and \cite{Me1, Paul2}, defining $N$ pairs of Grassmann fields, $\bar{c}^{r}$ and $c_{r}$, which transform in the (anti-)fundamental representation of $SU(N)$, giving them the following Poisson brackets:
\begin{equation}
	\{\bar{c}^{r}, c_{s}\}_{_{PB}} = -i\delta^r_{s}; \qquad \{\bar{c}^{r}, \bar{c}^{s}\}_{_{PB}} = 0 = \{c_{r}, c_{s}\}_{_{PB}}.
	\label{acomm}
\end{equation}
By taking $[T^a,T^b] =if^{abc}T^c$, it is easy to check that the expressions 
\begin{align}
R^{a} \equiv \bar{c}^{r} (T^{a})_{r}{}^sc_{s}~,
\end{align}
which can be used to absorb the gauge group indices of the generators, supply us with a (classical) representation of the Lie algebra, i.e.
\begin{equation}
	\left\{R^{a}, R^{b}\right\}_{_{PB}} = f^{abc}R^{c}~.
	\label{Lie}
\end{equation}
To correctly produce the above Poisson brackets between the auxiliary fields, their dynamics is specified as $S[\bar{c}, c] = \int d\tau \, i\bar{c}^{r} \dot{c}_{r}$. In the path integral formulation this first order action is responsible for producing the path-ordering prescription that is required to maintain gauge invariance for the coupling to a non-Abelian field. The full particle action thus reads
\begin{equation}
	S\left[\omega, p, \psi, e, \chi, \bar{c}, c \right] = \int_{0}^{1} d\tau \, \bigg[ p \cdot \dot{\omega} + \frac{i}{2}\psi \cdot \dot{\psi} +  i \bar{c}^{r} \dot{c}_{r} -e\widetilde{H} - i \chi \widetilde{Q}\bigg],
	\label{Sc}
\end{equation}
where  
\begin{equation}	
	\widetilde{H} = \widetilde{\pi}^{2} + \frac{i}{2} \psi^{\mu}F^{a}_{\mu\nu}\psi^{\nu} \bar{c}^{r}(T^{a})_r{}^{s} c_{s} ; \qquad \widetilde{Q} = \psi \cdot \widetilde{\pi}; \qquad \widetilde{\pi}^{\mu} = p^{\mu} - A^{a\mu} \bar{c}^{r}(T^{a})_r{}^{s}c_{s}
\end{equation}
and $F_{\mu\nu}^{a} = \partial_{\mu} A_{\nu}^{a} - \partial_{\nu}A_{\mu}^{a} + if^{abc}A_{\mu}^{b}A_{\nu}^{c}$ has been completed to the full (non-Abelian) field strength tensor. The above charges provide the modified constraints which impose the new mass-shell condition and the covariant Dirac equation $\gamma \cdot D \phy = 0$ in the presence of the gauge field. Due to (\ref{Lie}), the terms in $\widetilde{H}$ and $\widetilde{Q}$ which couple the gauge field to the particle worldline retain the correct group structure, while keeping the particle action gauge-invariant. This is easy to verify. By calling ${\cal U}$ a generic $SU(N)$ gauge transformation, the coloured fields transform as (colour indices are left implied when unnecessary)
\begin{align}
A_\mu\ \to\ {\cal U}\big(  A_\mu +i\partial_\mu \big) {\cal U}^\dagger,\quad c\ \to\ {\cal U} c,\quad \bar{c}\ \to\  \bar{c}~ {\cal U}^\dagger 
\end{align}  
and the momentum $\tilde \pi_\mu$ can be made gauge-invariant by imposing the gauge transformation
\begin{align}
p_\mu \ \to \ p_\mu -i\bar c~ {\cal U}^\dagger \partial_\mu {\cal U} c
\end{align}
that in turn makes the symplectic form $\int d\tau (p\cdot \dot\omega +i\bar c \dot{c})$ gauge-invariant. 
So the additional fields simply ensure the desired interactions are maintained without the need to rely on algebra valued potentials which would require path ordering upon functional quantisation. 

It is easy to see that there is an additional global $U(1)$ symmetry in (\ref{Sc}) which acts on the auxiliary Grassmann fields as $c_{r} \rightarrow e^{-i \vartheta}c_{r}$ and $\bar{c}^{r} \rightarrow \bar{c}^{r}e^{ i \vartheta}$. The conserved current associated to this symmetry is $L = \bar{c}^{r}c_{r}$ and it is useful to gauge this symmetry by modifying the action to
\begin{equation}
	S\left[\omega, p, \psi, e, \chi, \bar{c}, c, a \right] = \int_{0}^{1} d\tau \, \bigg[ p \cdot \dot{\omega} + \frac{i}{2}\psi \cdot \dot{\psi} + i\bar{c}^{r} \dot{c}_{r} - e\widetilde{H} - i \chi \widetilde{Q}-a(L - s)\bigg],
	\label{Sca}
\end{equation}
where the new gauge field $a(\tau)$ transforms under the $U(1)$ symmetry as $\delta a = \dot{\vartheta}$. We have also introduced a constant, $s = n - \frac{N}{2}$, to build a gauge invariant Chern-Simons term for the gauge field $S_{CS} = \int d\tau \, a(\tau) s$, which modifies the constraint produced by the equation of motion for $a(\tau)$ to $L + \frac{N}{2} = n$. To see why these quantities should be introduced we must examine the Hilbert space of the additional Grassmann fields. The anti-commutation relations associated to the canonical quantization of~\eqref{acomm} can be realised by promoting $\bar{c}^{r}$ and $c_{r}$ to creation and annihilation operators, $\hat c^{\dagger\, r}$ and $\hat{c}_{r}$, acting on coherent states
\begin{equation}
	\bstate{u} = \langle 0|e^{\bar u^{r} \hat{c}_{r}}; \qquad \bstate{u}\hat{c}^{\dagger r} = \bar{u}^{r}\bstate{u}; \qquad \bstate{u}\hat{c}_{r} = \partial_{\bar{u}^{r}}\bstate{u},
\end{equation}
whereby the operators naturally act by multiplication or derivation with respect to the Grassmann variables $\bar{u}^{r}$. Wave functions are then built out of components which transform as anti-symmetric tensor products of the original representation of the gauge field -- for a state $\state{\Psi}$ its associated wave function $\Psi(x, \bar{u}) = \left<\bar{u}, x|\Psi\right>$ has a finite expansion
\begin{equation}
	\Psi(x, \bar{u}) = \psi(x) + \psi_{r_{1}}(x)\bar{u}^{r_{1}} + \psi_{r_{1} r_{2}}(x) \bar{u}^{r_{1}}\bar{u}^{r_{2}} + \ldots + \psi_{r_{1} r_{2} \ldots  r_{N}}(x)\bar{u}^{r_{1}}\bar{u}^{r_{2}}\cdots\bar{u}^{r_{N}}~,
	\label{Psi}
\end{equation}
with totally antisymmetric tensors $\psi_{r_{1} r_{2} \ldots  r_{l}}$.
The wave function is consequently not described by an irreducible representation of the gauge group. However, the constraint function, $L$, becomes the number operator $\hat{L} = \hat{c}^{\dagger\, r} c_{r}$ whose eigenvalues indicate the occupation of each state. Acting on the Fock space we solve an operator ordering ambiguity by choosing the anti-symmetric combination $\hat{L} = \frac{1}{2}\left(\bar{u}^{r} \partial_{\bar{u}^{r}} - \partial_{\bar{u}^{r}} \bar{u}^{r}\right)$ so that the constraint imposed by the gauge field becomes
\begin{equation}
 	\left(\hat{L} + \frac{N}{2}\right)\state{\Psi} = n\state{\Psi} \longrightarrow \left(\bar{u}^{r} \frac{\partial}{\partial \bar{u}^{r}} - n\right)\Psi(x, \bar{u}) = 0.
\end{equation}
This constraint selects from (\ref{Psi}) the component that transforms with $n$ fully anti-symmetrised indices, $\psi_{r_{1}\ldots r_{n}}$, by enforcing all other components to vanish, therefore acting to project the wavefunction onto a single irreducible representation. This method has been used in the past to construct worldline theories describing higher spin fields \cite{Bastbrst, Bastvt}, differential forms \cite{Howe, Bastforms2} and one-loop gluon amplitudes in the presence of bosonic and spinor matter \cite{Bastwl1, Bastwl2}. 

There is an obvious limitation, however, in using the worldline action (\ref{Sca}) because the representations in which the matter field can  transform are restricted to those built out of fully antisymmetric tensor products of the original representation of the gauge group. In this work we generalise the worldline action to overcome this inadequacy and provide the means to project onto any, arbitrarily chosen representation of the symmetry group. This provides a complete framework in which the worldline approach can be applied to any form of fermionic matter field by merely introducing some additional degrees of freedom to take part in quantisation whose r\^{o}le is to ensure that the contribution from undesired wavefunction components are excluded from physical results. 

\section{Extension of the worldline action}
\label{SecExtension}
To achieve a richer Hilbert space associated to the additional Grassmann fields the simple generalisation we propose is to include $F$ families of these ``colour'' fields. We will denote the families by an additional subscript (continuing to prescribe to the summation convention) and define the generalised Poisson brackets
\begin{equation}
	\{\bar{c}_{f}^r, c_{gs}\}_{_{PB}} = -i\delta_{fg}\delta^r_{s}; \qquad \{\bar{c}_{f}^r, \bar{c}_{g}^s\}_{_{PB}} = 0 = \{c_{fr}, c_{gs}\}_{_{PB}}.
	\label{acommf}
\end{equation}
The worldline action (\ref{Sc}) is modified to 
\begin{equation}
	S\left[\omega, p, \psi, e, \chi, \bar{c}, c\right] = \int_{0}^{1} d\tau \, \bigg[ p \cdot \dot{\omega} + \frac{i}{2}\psi \cdot \dot{\psi} + i \bar{c}_{f}^r \dot{c}_{rf} - e\widetilde{H} - i \chi \widetilde{Q}\bigg],
	\label{Scf}
\end{equation}
where in the tilded quantities the conjugate momentum now involves a sum over the $F$ families: $\widetilde{\pi}_{\mu} = p^{\mu} - A^{a}_\mu\bar{c}_{f}^r(T^{a})_{r}{}^sc_{fs}$. This action remains invariant under local supersymmetry transformations generated by $\widetilde{Q}$ and local reparameterisations generated by $\widetilde{H}$, but the global $U(1)$ symmetry that was previously encountered has been enlarged to a non-Abelian ``flavour group'' $U(F)$. Specifically, if $\Lambda= e^{-i\lambda}$ is an element of the global symmetry group,
the  action (\ref{Scf}) is invariant under $c_{fr} \rightarrow \Lambda_{fg} c_{gr}$ and $\bar{c}_{f}^r \rightarrow \bar{c}_{g}^r \Lambda^\dagger_{gf}$.

We find the conserved currents associated to these transformations by using the Noether trick. The result is a generalisation of the occupation number\footnote{Note that setting $g = f$, then $L_{ff}$ coincides with the familiar current related to the occupation number of family $f$.} $L_{fg} = \bar{c}_{f}^rc_{fg}$. These currents satisfy the following $U(F)$ algebra
\begin{equation}
	\left\{L_{fg}, L_{ f^{\prime}g^{\prime}}\right\}_{_{PB}} = i\delta_{fg'}L_{f'g} - i\delta_{f'g}L_{fg'}
\end{equation}
so they can be used to build an infinitesimal $U(F)$ transformation by taking Poisson brackets with $\lambda = \lambda_{fg}L_{fg}$:
\begin{equation}
	\delta_{\lambda}c_{fr} = \left\{c_{fr}, \lambda\right\}_{_{PB}} = -i\lambda_{fg}c_{gr}; \qquad \delta_{\lambda}\bar{c}_{f}^r = \left\{\bar{c}^{r}_{f}, \lambda\right\}_{_{PB}} = i\bar{c}_{g}^r\lambda_{gf}.
	\label{deltaUf}
\end{equation}
The Fock space associated to the Grassmann fields is also more complicated, because one can now act on the vacuum with creation operators associated to different families:
\begin{equation}
	\Psi(x, \bar{u}) = \sum_{\{n_{1}, n_{2}, \ldots n_{\!F}\}} \psi_{r_{1} \ldots r_{n_{\!1}},s_{1} \ldots s_{n_{2}},\ldots ,t_{1} \ldots t_{n_{F}}}(x)~\bar{u}_{\!F}^{t_{1}}\!...\bar{u}_{F}^{t_{n_{\!F}}}\cdots\bar{u}_{2}^{s_{1}}\!\!...\bar{u}_{2}^{s_{n_{\!2}}}\bar{u}_{1}^{r_{1}}\!\!...\bar{u}_{1}^{r_{n_{\!1}}}
	\label{Psif}
\end{equation}
where each component is completely antisymmetric in the blocks of indices associated to the individual families (but has no special symmetry between indices from separate families). So wavefunctions are in general described by a reducible representation of the symmetry group, with the components in (\ref{Psi}) transforming as tensor products of the representation created by each family -- affecting a Young-Tableaux notation we may write
\begin{equation}
\Yvcentermath1
	\Psi(x, \bar{u}) \sim \sum_{\{n_{1}, n_{2}, \ldots n_{\!F}\}}  \underbrace{\underset{ \,\Small \yng(1,1)\, }{\overset{ \,\Small\yng(1,1)\, } {\rvdots} }}_{n_{F}} \otimes  \ldots \otimes \underbrace{\underset{ \,\Small\yng(1,1)\, }{\overset{ \,\Small\yng(1,1)\, } {\rvdots} }}_{n_{2}}   \otimes \underbrace{\underset{ \,\Small\yng(1,1)\, }{\overset{ \,\Small\yng(1,1)\, } {\rvdots} }}_{n_{1}}   
	\label{PsiTab}
	\Yvcentermath0
\end{equation}
to highlight how the components transform (the $f^{th}$ term in the product has $n_{f}$ boxes, representing $n_{f}$ fully anti-symmetrised indices). We have made a positive step in terms of enlarging the Hilbert space to include components transforming in more general representations, but it is now necessary to generalise the procedure which projects out undesired contributions to physical phenomena in order to select a chosen component that transforms irreducibly from (\ref{PsiTab}).

\subsection{Gauging the $U(F)$ symmetry}
To attain the desired projection we follow the ideas of Section \ref{SecTheory} by gauging the $U(F)$ symmetry (\ref{deltaUf}) through the introduction of gauge fields $a_{fg}(\tau)$. Under a local $U(F)$ transformation $a_{fg}(\tau)$ is taken to transform as
\begin{equation}
	\delta_{\lambda} a_{fg}(\tau) = \dot{\lambda}_{fg}(\tau) + i\left[a(\tau),\lambda(\tau)\right]_{fg}
	\label{deltaa}
\end{equation}
which ensures the invariance of 
\begin{equation}
	S\left[\omega, p, \psi, e, \chi, \bar{c}, c, a\right] = \int_{0}^{1} d\tau \, \bigg[ p \cdot \dot{\omega} + \frac{i}{2}\psi \cdot \dot{\psi} + i \bar{c}_{f}^r \dot{c}_{rf} - e\widetilde{H} - i \chi \widetilde{Q} -a_{fg}L_{fg} \bigg],
	\label{Scaf}
\end{equation}
under local $U(F)$ transformations. At this juncture, we would also like to introduce an appropriate Chern-Simons term for each of the diagonal entries $L_{f} \equiv L_{ff}$ (no sum implied), so as to project onto the subspace of components which have a given number of indices associated with each family. This means that it is not possible to gauge the entire $U(F)$ group, since in that case the only Abelian invariant\footnote{Note that $\{L_{f}, L_{f^{\prime}}\}_{_{PB}} = 0$ but $\{L_{f}, L_{gg^{\prime}}\}_{_{PB}} = i\delta_{fg}L_{fg^{\prime}} - i\delta_{fg^{\prime}}L_{gf}$ does not vanish in general. Summing over $f$, however, is sufficient to produce an invariant $U(1)$ generator $L$.} is $L = \sum_{f}L_{f}$. This would only allow the addition of a single Chern-Simons term $s \int d\tau \sum_{f}  a_{ff}$ and we would be unable to control the occupation number of individual families separately. Instead, we begin by gauging the $U(1)^{F} \subset U(F)$ subgroup of the full symmetry group by introducing the diagonal fields $a_{f} \equiv a_{ff}$ and diagonal generators $L_{f}$ only. In this case the $F$ diagonal gauge fields $a_{f}$ have an Abelian transformation. We can thus introduce $F$ Chern-Simons terms so as to impose the constraints on the Fock space
\begin{equation}
	\left(\hat{L}_{f} + \frac{N}{2}\right)\state{\Psi} = n_{f}\state{\Psi} \longrightarrow \left(\bar{u}^{r}_{f}\frac{\partial}{\partial \bar{u}_{f}^{r}} - n_{f}\right)\Psi(x, \bar{u}) = 0,
	\label{consF}
\end{equation}
which is sufficient to project onto the wavefunction component with $n_{f}$ antisymmetric indices associated to each family, $\psi_{r_{1} \ldots r_{n_{\!1}},s_{1} \ldots s_{n_{\!2}},\ldots, t_{1} \ldots t_{n_{\!F}}}(x)$.

Gauging only $U(1)^{F}$ is not sufficient to achieve irreducibility, since the selected wavefunction component itself transforms as a tensor product of the antisymmetric representation created by each family (see (\ref{PsiTab}) for illustration). To then pick out a single irreducible representation we must gauge a larger part of the $U(F)$ symmetry whilst keeping this subgroup invariant. Consider the partial gauging of those $L_{fg}$ with $1 \leqslant g \leqslant f \leqslant F$, for which the following first class sub-algebra holds
\begin{alignat}{2}
	\left\{L_{fg}, L_{f^{\prime}f\phantom{^{\prime}}}\right\}_{_{PB}} &= -i\delta_{f'g}L_{f} + iL_{f'g}  \qquad & g^{\prime} = f \nonumber \\
	\left\{L_{fg}, L_{f^{\prime}g^{\prime}}\right\}_{_{PB}} &= -i\delta_{f^{\prime}g}L_{fg^{\prime}}  \qquad  & g^{\prime} < f  \nonumber \\
	\left\{L_{fg}, L_{f^{\prime}g^{\prime}}\right\}_{_{PB}} &= 0  \qquad & f < g^{\prime} .
\end{alignat} 
So the subalgebra is first class. Later we will refer to the group associated to these generators as the ``auxiliary gauge group.'' With this choice it is easy to check that the diagonal elements of the gauge field still transform in an Abelian manner,
\begin{align}
	\delta_{\lambda} a_{fg}(\tau) &= \dot{\lambda}_{fg}(\tau) + i\sum_{g \leqslant k \leqslant f}\left(a_{fk}(\tau)\lambda_{kg}(\tau)-\lambda_{fk}(\tau)a_{kg}(\tau)\right) \label{deltaag} \\
	\Rightarrow \delta_{\lambda}a_{ff}(\tau) &= \dot{\lambda}_{ff}(\tau),		
\end{align}
which means that $F$ independent Chern-Simons terms remain admissible and the constraints (\ref{consF}) continue to be imposed. We therefore arrive at the complete expression for the worldline action in phase-space
\begin{align}
	S^{\prime}\left[\omega, p, \psi, e, \chi, \bar{c}, c, a\right]\! =\! \int_{0}^{1}\! d\tau \bigg[ p \cdot \dot{\omega} &+ \frac{i}{2}\psi \cdot \dot{\psi} + i\bar{c}_{f}^r \dot{c}_{rf} - e\widetilde{H} - i \chi \widetilde{Q}  \nonumber \\
	&-\textstyle{\sum\limits_{f=1}^{F}}a_{f}\left(L_{f} - s_f\right)- \textstyle{\sum\limits_{g < f}}a_{fg}L_{fg} \bigg],
	\label{Smain}
\end{align}
where we have used the notation $a_{f} \equiv a_{ff}$. 
This is the worldline theory that we will use for the rest of this paper. First we examine the effect that the larger gauging has on the Fock space of the Grassmann fields and in the next section we will quantise the worldline theory in the path integral formulation. 

In canonical quantisation the $U(F)$ generators $L_{fg}$ become (for $g < f$) $\hat L_{fg} = \bar{u}^{r}_{f} \partial_{\bar{u}^{r}_{g}}$ when acting on coherent states and the equation of motion for $a_{fg}$ imposes the constraint 
\begin{equation}
	\hat{L}_{fg} \state{\Psi} = 0 ~\longrightarrow ~\bar{u}^{r}_{f}\frac{\partial}{\partial \bar{u}^{r}_{g}} \Psi(x, \bar{u}) = 0.
	\label{Lfg}
\end{equation}
To understand how this additional constraint acts on the wavefunction selected by the $L_{f}$, it suffices to recall that $\psi_{r_{1} \ldots r_{n_{\!1}}, s_{1} \ldots s_{n_{2}},\ldots, t_{1} \ldots t_{n_{\!F}}}(x)$ can be written as a sum over components transforming in irreducible representations with various symmetries between indices belonging to the different families. For example, taking $F = 2$, and $n_{2} = 2$, $n_{1} = 1$, the surviving part -- satisfying (\ref{consF}) -- of $\Psi(x, \bar{u})$ can be written (exploiting the $S_{F}$ symmetry allowing the permutation of the $n_{f}$) as
\begin{align}
	\Psi(x, \bar{u}) &= \psi_{r_{1} r_{2},s_{1}}(x) \,\bar{u}^{r_{1}}_{2}\bar{u}^{r_{2}}_{2}\bar{u}^{s_{1}}_{1} \nonumber \\
	&= \left[\psi_{[r_{1}r_{2},s_{1}]}(x) + \frac{1}{3}\big( 2\psi_{r_{1}r_{2},s_{1}}(x) - \psi_{s_{1}r_{1},r_{2}}(x) - \psi_{r_{2}s_{1},r_{1}}(x) \big)\right]\bar{u}^{r_{1}}_{2}\bar{u}^{r_{2}}_{2}\bar{u}^{s_{1}}_{1},
	\label{Psidec}
\end{align}
which represents the tensor product decomposition $\Yvcentermath1{\Tiny \yng(1,1)} \otimes {\Tiny \yng(1)} = {\Tiny \yng(1,1,1) } \oplus {\Tiny \yng(2,1)} \Yvcentermath0$. The constraints (\ref{Lfg}) imply that only the components in the kernel of all the $\hat{L}_{fg}$ survive. Each $\hat{L}_{fg}$ acts as an ``anti-symmetriser'', i.e. it replaces a $\bar{u}_{g}^{r}$ with a $\bar{u}_{f}^{r}$ and anti-symmetrises over the indices involved. For the current example, the only non-diagonal number operator that exists is $\hat{L}_{21}$, which acts as
\begin{align}
	\hat{L}_{21}\,\bar{u}^{r_{1}}_{2}\bar{u}^{r_{2}}_{2}\bar{u}^{s_{1}}_{1} &= \bar{u}_{2}^{t}\frac{\partial}{\partial \bar{u}^{t}_{1}}\,\bar{u}^{r_{1}}_{2}\bar{u}^{r_{2}}_{2}\bar{u}^{s_{1}}_{1} \nonumber \\
	&=\bar{u}^{s_{1}}_{2} \bar{u}^{r_{1}}_{2}\bar{u}^{r_{2}}_{2}
\end{align}
which is now anti-symmetric under interchange of the indices $r_{1}$, $r_{2}$ and $s_{1}$. It is easy to verify that the result of contracting the indices of the term in rounded brackets of (\ref{Psidec}) with the above product identically vanishes (this component transforms in the $ \Yvcentermath1 {\Tiny \yng(2,1)}  \Yvcentermath0$ representation which implies symmetrisations over the $r_{1}$, $s_{1}$ and $r_{2}$, $s_{1}$ pairs). This means that the constraint acts as
\begin{equation}
	0=\hat{L}_{21} \Psi(x, \bar{u})  = \psi_{[r_{1}r_{2},s_{1}]} (x) \bar{u}^{r_{1}}_{2}\bar{u}^{r_{2}}_{2}\bar{u}^{s_{1}}_{2}\ \Longrightarrow\ \psi_{[r_{1}r_{2},s_{1}]} (x)=0
	\label{proj2}
\end{equation}
so that the only surviving component of the wavefunction is that which transforms in the $ \Yvcentermath1 {\Tiny \yng(2,1)}  \Yvcentermath0$ representation. This is described by a Young Tableau with $n_{2}$ rows in the first column and $n_{1}$ rows in the second column.

In general, the effect of the constraints can be summarised as follows. The diagonal operators $\hat{L}_{f}$ act to select the occupation number of each family of Grassmann fields to be equal to $n_{f}$, picking out a single wavefunction component (i.e. with fixed $n_{f}$) from the tensor product (\ref{PsiTab}). The off-diagonal operators $\hat{L}_{fg}$ then select from this reducible sum a single component with compatible symmetries. To lie in the kernel of all of the $\hat{L}_{fg}$ requires the maximum inter-family symmetry so that the single surviving component that the constraints project onto transforms in the representation with Young Tableau
\begin{equation}
\Yvcentermath1
	\Psi(x, \bar{u}) \sim  \underbrace{\overset{n_{F}...\hfill}{\underset{ \Small \yng(4,2,1) }{ \overset{ \Small \yng(4,4,4) }{ \rvdots } }} \overset{\overset{...}{\vphantom{ { \yng(1)} }}}{\underset{ \underset{...}{ \vphantom{  {\yng(1)}  } } }{...}}   \overset{\hfill ...n_{1}} {\vphantom{\underset{ \Small \yng(4,2,1) }{ \overset{ \Small \yng(4,4,4) }{ \rvdots } }} \underset{ \Small \yng(2) \hphantom{\yng(2)}  }{ \overset{ \Small \yng(4,3,2) }{ \rvdots } }}}_{F \textrm{ columns}}
\Yvcentermath0
\label{Psiirrep}
\end{equation}
which has $n_{F}$ rows in the first column, $n_{F-1}$ rows in the second and so forth, where $N \geqslant n_{F} \geqslant n_{F-1} \geqslant \ldots \geqslant n_{2} \geqslant n_{1}$. The symmetrisation between indices associated to different families implied by (\ref{Psiirrep}) means that this is the only component that lies in the kernel of all of the $\hat{L}_{fg}$. This ensures that the wavefunction transforms in an irreducible representation of $SU(N)$. The conclusion is that with the current approach it is possible to project onto an arbitrary representation, without being restricted to those with single-column Tableaux. We provide a much more versatile method for the worldline description of $SU(N)$-charged matter fields forming any multiplet. 

In the next section we verify this analysis by carrying out functional quantisation for a particle traversing a closed path. This has application in the worldline formalism where it is related to the partition function and one-loop effective action of the quantum field theory describing matter interacting with a gauge field. We will do so by counting the number of degrees of freedom (DoF) associated to the matter field for comparison with the dimension of the representation denoted by (\ref{Psiirrep}).

\section{Path integral quantisation}
\label{SecPath}
To facilitate the path integral quantisation of the particle described by (\ref{Smain}) we first rewrite the action in configuration space by solving the equation of motion for $p^{\mu}$ and putting it on-shell. This leads to a (Euclidean) path integral 
\begin{equation}
	\int 	\frac{\mathscr{D}e \mathscr{D} \chi \mathscr{D}\omega \mathscr{D}\psi \mathscr{D}\bar{c}\mathscr{D}c\mathscr{D}a}{\rm Vol(Gauge)} \, \exp{\left(-S\left[\omega, \psi, e, \chi, \bar{c}, c, a\right]\right)}
	\label{Path}
\end{equation}
where 
\begin{align}
	S\left[\omega, \psi, e, \chi, \bar{c}, c, a\right] =  \int_{0}^{1} d\tau \bigg[e^{-1}\frac{\dot{\omega}^{2}}{2} + \frac{1}{2}\psi  \cdot \dot{\psi} - \frac{\chi}{e} \dot{\omega} \cdot \psi + \bar{c}^{r}_{f} \dot{c}_{fr} &- i\bar{c}^{r}_{f} \mathcal{A}^{a}(T^{a})_{r}{}^{s} c_{fs}  \nonumber \\
	+ \textstyle{\sum\limits_{f=1}^{F}}ia_{f}\left(\bar{c}^{r}_{f}c_{fr}\! - \!s_{f}\right) &+ \textstyle{\sum\limits_{g < f}}ia_{fg}\bar{c}^{r}_{f}c_{gr} \bigg].
	\label{Smainc}
\end{align}
Above, the division by the volume of the gauge group (i.e. diffeomorphisms, supersymmetry and auxiliary gauge group) takes care of summing only over gauge-inequivalent configurations.
In equation~\eqref{Smainc} we have denoted $\mathcal{A} = \dot{\omega} \cdot A - \frac{e}{2}\psi^{\mu}F_{\mu\nu}\psi^{\nu}$ which makes up the Wilson loop exponent as $W = \mathscr{P} e^{i \int_{0}^{1} \mathcal{A}(\tau) d\tau}$ and we denote the $F$ independent Chern-Simons levels by $s_{f}$. The path ordering prescription is now taken care of by the inclusion of the additional Grassmann fields. The ends of the interval are identified to form $S^{1}$ -- the boundary conditions on $\omega$, $e$ and $a$ are periodic, whereas the Grassmann fields $\psi$, $\chi$, $\bar{c}$ and $c$ are taken to be anti-periodic. This form of the action would be crucial in the context of the worldline formalism where it would form the basis for the calculation of multi-loop scattering amplitudes. Indeed, the quantisation of the $N = 1$ point particle leads to the functional integrals involving the embedding coordinates, $\omega$, its spin degree of freedom, $\psi$ and the worldline supergravity multiplet $e$ and $\chi$. For simplicity, however, we will focus here only on the dimension associated to the colour degrees of freedom. This allows us to focus on the restricted path integral relating to the auxiliary Grassmann fields and to temporarily drop the coupling to the external gauge field (so we set $A = 0$). The restricted path integral which counts the particle degrees of freedom is then
\begin{align}
	\frac{1}{\rm Vol(Gauge)}\int &{\prod_{f,r}\mathscr{D}\bar{c}^{r}_{f}\mathscr{D}c_{fr} \prod_{h < g}\mathscr{D}a_{gh} \prod_f\mathscr{D}a_f}\nonumber \\
	&\times \exp{\left(-\int_{0}^{1} d\tau \, \left[  \textstyle{\sum\limits_{f=1}^{F}} \left(\bar{c}^{r}_{f} D_{f} c_{fr} - is_{f}a_{f}\right) +  \textstyle{\sum\limits_{h < g}}ia_{gh}\bar{c}^{r}_{f}c_{gr} \right]\right)},
	\label{pic}
\end{align}
we we have used the diagonal gauge fields to construct the covariant derivative $D_{f} \equiv \left(\frac{d}{d\tau} + ia_{f}\right)$, and ``Vol(Gauge)'' now only refers to the auxiliary gauge group. The calculation of this functional integral will be shown to yield the correct number of degrees of freedom associated to the colour space for any choice of $s = \left(s_{1}, s_{2}, \ldots, s_{F}\right)$. 

The action in (\ref{pic}) enjoys the (local) auxiliary gauge group symmetry that was discussed in the previous section and so to compute the path integral we must first employ a gauge fixing procedure. On the circle, it is not possible to entirely gauge away the fields $a_{fg}$; at best one may set the $a_{kj}$ to be constant (see Appendix~\ref{sec:appA}). ~The most general form of $a_{kj}$ would be a constant upper-triangular matrix, however as discussed in Appendix~\ref{sec:appA} the functional integral over the colour fields is independent of the off-diagonal entries so it suffices to take
\begin{equation}
	\hat{a}_{fg} = \begin{pmatrix}
		\theta_{1} & 0 & \cdot & 0 \\ 0 & \theta_{2} & \cdot & 0 \\ \cdot & \cdot & \cdot & \cdot \\ 0 & 0 & \cdot & \theta_{F}
	\end{pmatrix}.
	\label{afix}
\end{equation}
The $\{\theta_{f}\}$ then remain as moduli to be integrated over and can be identified as angles in $[0, 2\pi]$ by examining large $U(1)^{F}$ transformations. This gauge fixing can be compensated for by introducing Faddeev-Popov ghosts, $\gamma_{fg}$ and $\bar{\gamma}_{fg}$, which ensure that gauge invariance is maintained. The precise form of the ghost action depends on how the global $U(F)$ symmetry was gauged. With the partial gauging discussed in the previous section, the gauge transformation of the $\hat{a}_{fg}$ is given by (\ref{deltaag}):
\begin{align}
	\delta  \hat{a}_{gh} &= \dot{\lambda}_{gh} + i \sum_{h \leqslant k \leqslant g}\left(\theta_{g}\delta_{gk}\lambda_{kh} - \theta_{h}\lambda_{gk}\delta_{kh}\right) \nonumber \\
	&= \left(\frac{d}{d\tau} + i\left(\theta_{g} - \theta_{h}\right)\right)\lambda_{gh} 
\end{align}
which leads to a ghost action
\begin{equation}
	S_{g}\left[\bar{\gamma}, \gamma, \{\theta\}\right] = \int_{0}^{1}d\tau\,  \textstyle{\sum\limits_{h \leqslant g}}\, \bar{\gamma}_{gh}\left(\frac{d}{dt} + i\left(\theta_{g} - \theta_{h}\right)\right)\gamma_{gh}.
	\label{Sghost}
\end{equation}
The result of these calculations is the following gauge fixed path integral
\begin{align}
	\prod_{f = 1}^{F} \int _{0}^{2\pi} \frac{d\theta_{f}}{2\pi} \int \mathscr{D} (\bar{c},c) \mathscr{D}(\bar{\gamma},\gamma) \, \exp\bigg(&- \int_{0}^{1}\!d\tau\bigg[ \sum_
{f = 1}^{F} \left( \bar{c}^{r}_{f} D_{f} c_{fr} - i\theta_{f}s_{f} \right) \nonumber \\
	&\qquad\qquad \qquad+\sum_{h \leqslant g} \bar{\gamma}_{gh}\left(\frac{d}{dt}\! + \!i\left(\theta_{g}\! -\! \theta_{h}\right)\right) \gamma_{gh}\bigg] \bigg)
\end{align}
where on the chosen gauge slice $D_{f} \rightarrow \left(\frac{d}{d\tau} + i \theta_{f}\right)$ and the $s_{f}$ are determined by our choice of population numbers for each family of colour fields. The ghosts are taken to be periodic around the circle. The functional integration leads to a product of various determinants which we now calculate.

\subsection{Evaluation of the functional integrals}
The integration over the colour degrees of freedom factorises due to our choice of gauge fixing the fields of the auxiliary gauge group to $\hat{a}_{fg}$ and provides a product of functional determinants
\begin{equation}
	\prod_{f = 1}^{F}\left[\underset{\textrm{\Tiny \!\!APB}}{\Det}\left(\frac{d}{d\tau} + i\theta_{f}\right)\right]^{N}
\end{equation}
which we define as the product of their eigenvalues for functions on $S^{1}$ with anti-periodic boundary conditions (APB). In Appendix B we show that this yields
\begin{equation}
	\prod_{f = 1}^{F} \left(2\cos{\left(\frac{\theta_{f}}{2}\right)}\right)^{N}.
\end{equation}
Similarly, integrating over the ghosts leads to a product of functional determinants on functions with periodic boundary conditions (PB), defined as the product of their non-zero eigenvalues, which effectively provides a measure for the $U(F)$ moduli:
\begin{equation}
	\mu\left(\{\theta_{k}\}\right) = \prod_{f=1}^{F}\underset{\textrm{\Tiny \!\!PB}}{\Det}\left(\frac{d}{d\tau}\right)\times\prod_{h \leqslant g}\underset{\textrm{\Tiny \!\!PB}}{\Det}\left(\frac{d}{d\tau} + i \left(\theta_{g} - \theta_{h}\right)\right)
\end{equation}
which evaluates to (see Appendix B)
\begin{equation}
	\mu\left(\{\theta_{k}\}\right) = \prod_{h < g}2i \sin{\left(\frac{\theta_{g} - \theta_{h}}{2}\right)}.
	\label{mu}
\end{equation}
Putting this together we find that the number of colour degrees of freedom is represented by a simple multiple integral
\begin{equation}
	K_{F}\prod_{f = 1}^{F} \int _{0}^{2\pi} \frac{d\theta_{f}}{2\pi}e^{i \theta_{f}s_{f}} \mu\left(\{\theta_{k}\}\right) \left(2\cos{\left(\frac{\theta_{f}}{2}\right)}\right)^{N}
	\label{main}
\end{equation}
which represents the main result of this article. In this formula, $K_{F}$ is a normalisation constant whose inverse is equal to the number of fundamental domains included in the integration over the $U(F)$ moduli.

We illustrate the application of this formula in a few special cases, before rewriting it in terms of more convenient variables. We first verify that for $F = 1$ formula (\ref{main}) reduces to the previously calculated result for totally anti-symmetric representations. Then we demonstrate the case $F = 2$ to prove that the correct counting of DoF is reached for the example discussed in (\ref{Psidec}). So starting with only one family of additional Grassmann fields we take $s_{1} = n_{1}-\frac{N}{2}$ and note that the measure, $\mu\left(\theta_{1}\right) = 1$, is trivial and $K_{1} = 1$. Then (\ref{main}) becomes
\begin{align}
	&\int_{0}^{2\pi} \frac{d\theta_{1}}{2\pi} e^{i\theta_{1} n_{1}}e^{-i\theta_{1}\frac{N}{2}}\left(e^{i \frac{\theta_{1}}{2}} + e^{-i \frac{\theta_{1}}{2}}\right)^{N} \nonumber \\
	=&\int_{0}^{2\pi} \frac{d\theta_{1}}{2\pi} e^{i\theta_{1} n_{1}}\left(1 + e^{-i\theta_{1}}\right)^{N}.
\end{align} 
At this stage it is convenient to effect a change of variable by noting that $z_{1} = e^{i\theta_{1}}$ parametrises the $U(1)$ Wilson-loop on the circle. This leads to
\begin{align}
	&\oint \frac{dz_{1}}{2\pi i } z_{1}^{n_{1}-1}\left(1 + \frac{1}{z_{1}}\right)^{N} \nonumber \\
	=&\oint \frac{dz_{1}}{2\pi i}\frac{\left(z_{1} + 1\right)^{N}}{z_{1}^{N + 1 - n_{1}}},
\end{align}
where the contour of integration is taken to be the circle $\left|z_{1}\right| = 1$. For $n_{1} \leqslant N$ there is a pole at $z = 0$ and we can arrive at the elementary result
\begin{equation}
	\oint \frac{dz_{1}}{2\pi i}\frac{\left(z_{1} + 1\right)^{N}}{z^{N + 1 - n_{1}}} = \begin{cases}\nCr{N}{n_{1}} & 0 \leqslant n_{1} \leqslant N \\ ~~~0 & \textrm{Otherwise}\end{cases}.
	\label{contour}
\end{equation}
This shows that for $F = 1$, the main formula (\ref{main}) correctly counts the colour degrees of freedom for a matter field transforming in the representation with $n_{1}$ fully anti-symmetrised indices (this corresponds to the dimension of the single-column Young Tableau with $n_{1}$ rows). This result is in agreement with \cite{Bastwl1, Bastwl2}, where scattering amplitudes were calculated using worldline techniques for matter fields transforming in representations built out of antisymmetric tensor products of the fundamental representation of $SU(N)$. 

A less trivial example is that corresponding to (\ref{Psidec}). It is useful to consider this calculation in order to see how the modular measure provides the required projection onto an irreducible representation. To describe such a wavefunction we need $F = 2$ families of the additional Grassmann fields and for generality will take occupation numbers $n = (n_{1}, n_{2})$ with $0 \leqslant n_{1} \leqslant n_{2} \leqslant N$. To achieve this requires the Chern-Simons levels to be chosen as $s_{1} = n_{1} - \frac{N}{2} - \frac{1}{2}$ and $s_{2} = n_{2}-\frac{N}{2} + \frac{1}{2}$. Then application of (\ref{main}) provides ($K_{2}= 1$)
\begin{equation}
	\int_{0}^{2\pi}\frac{d\theta_{1}}{2\pi}\int_{0}^{2\pi}\frac{d\theta_{2}}{2\pi} e^{i \theta_{1}\left(n_{1}-\frac{1}{2}\right)}e^{i \theta_{2}\left(n_{2}+\frac{1}{2}\right)}\left(1 + e^{-i\theta_{1}}\right)^{N}\left(1 + e^{-i\theta_{2}}\right)^{N}\left(e^{i \frac{\theta_{1}}{2}}e^{-i\frac{\theta_{2}}{2}} - e^{-i \frac{\theta_{1}}{2}}e^{i\frac{\theta_{2}}{2}}\right)
\end{equation}
where we have already cancelled the terms $e^{-i \theta_{1} \frac{N}{2}}$ and $e^{-i\theta_{2}\frac{N}{2}}$ by factorising their inverses from the first two terms in rounded brackets. It is again convenient to make use of the Wilson loop variables $z_{1} = e^{i\theta_{1}}$ and $z_{2} = e^{i\theta_{2}}$ to rewrite this expression as an integration on the complex plane:
\begin{equation}
	\oint\!\!\oint  \frac{d z_{1}}{2\pi i}\frac{d z_{2}}{2\pi i} \left(\frac{\left(z_{1} + 1\right)^{N}}{z_{1}^{N+1-n_{1}}}\frac{\left(z_{2} + 1\right)^{N}}{z_{2}^{N+1 - n_{2}}} - \frac{\left(z_{1} + 1\right)^{N}}{z_{1}^{N + 1 - (n_{1} - 1)}}\frac{\left(z_{2} + 1\right)^{N}}{z_{2}^{N + 1 - (n_{2} + 1)}}\right).
\end{equation}
Using (\ref{contour}) this is easily determined to be
\begin{align}
	&\nCr{N}{n_{1}}\nCr{N}{n_{2}} - \nCr{N}{n_{1} - 1}\nCr{N}{n_{2} + 1} \nonumber \\
	=&\, \frac{N!}{(N - n_{2})!}\frac{(N+1)!}{(N - (n_{1} - 1))!}\frac{n_{2} + 1 - n_{1}}{n_{1}!(n_{2} + 1)!} \nonumber \\
	=&\,  \nCr{N}{n_{2}}\nCr{N\!+\!1}{n_{1}} \frac{n_{2} + 1 - n_{1}}{n_{2} + 1} \,.
	\label{dof2}
\end{align}
It is straightforward to confirm that this is the dimension of the representation with Young Tableau
\begin{equation}
	\Yvcentermath1
	\overset{n_{2}\,n_{1}}{\underset{ \Small \yng(2,1,1) }{ \overset{ \Small \yng(2,2,2) }{ \rvdots } }}
	\Yvcentermath0
\end{equation}
having $n_{2}$ rows in the first column and $n_{1}$ rows in the second\footnote{This diagram has ``factors'' $\frac{N!}{(N - n_{2})!}\frac{(N+1)!}{(N - (n_{1} - 1))!}$ and ``hooks'' $n_{1}!(n_{2} - n_{1})!\frac{(n_{2} + 1)!}{(n_{2} + 1 - n_{1})!}$ which divide to give its dimension in agreement with (\ref{dof2}) -- see \cite{Georgi}.}. In particular, setting $n_{2} = 2$ and $n_{1} = 1$ projects onto the representation $\Yvcentermath1 {\Tiny \yng(2,1)}\Yvcentermath0$ as in (\ref{proj2}). 

Returning now to the general case it is useful to incorporate the approach taken in these two examples by making a complete change of variables in (\ref{main}) by introducing $F$ complex variables $z_{f} = e^{i\theta_{f}}$ and to present the main formula in terms of these Wilson loop variables. In general, to project onto the $SU(N)$ representation specified by occupation numbers $n = \left(n_{1}, n_{2}, \ldots n_{F}\right)$ requires Chern-Simons levels $s_{f} = n_{f} - \frac{N}{2} - \frac{F - \left(2f-1\right)}{2}$ to impose the correct constraints on the physical state space. Overall, following some elementary manipulations the general formula can be written rather concisely in a similar manner to the examples above as
\begin{equation}
	K_{F}\prod_{f = 1}^{F} \oint \frac{d z_{f}}{2\pi i}\, \prod_{h < g}\left(1 - \frac{z_{h}}{z_{g}}\right) \frac{\left(z_{f} + 1\right)^{N}}{z_{f}^{N + 1 - n_{f}}},
	\label{main2}
\end{equation}
which represents the colour degrees of freedom for the matter field in a first quantised approach. The complex integrals simply pick out the poles at $z_{f} = 0$ for each term in the expansion of the product in rounded brackets. We have checked that this provides the correct counting of degrees of freedom for arbitrary representations of $SU(N)$ given by Young Tableau with $n_{f}$ rows in the $f^{th}$ column. The final formula (\ref{main2}) is applied by fixing the vector $n$ and expanding the expression into a sum over products of complex integrals. For illustration, setting $F = 6$ and $n = (1, 3, 5, 6, 6, 8)$, assuming $N \geqslant 8$, the integration evaluates to
\begin{equation}
\Yvcentermath1
	\textrm{dim} ~ {\Small \yng(6,5,5,4,4,3,1,1)}
\Yvcentermath0
\end{equation}
which shows that the partial gauging of the $U(F)$ symmetry group correctly projects onto the desired sector in the Hilbert space of the colour fields. Returning to (\ref{Path}), the remaining functional integration over the worldline matter and super-gravity fields generates the partition function for the dynamics and spin degrees of freedom of the particle. In the simple case we have considered here (without turning on the gauge field) we just reproduce the well known functional quantisation of a free point particle multiplet. We leave it to future work to demonstrate that, when the gauge field is coupled to the particle as in (\ref{Path}), the colour degrees of freedom produce the correct Wilson-loop interaction for the chosen representation of $SU(N)$. In that context, one of the physically significant representations that could be used is the adjoint of $SU(N)$. This requires the use of $F = 2$ families and the choice $n = (1, N-1)$, whereby (\ref{dof2}) evaluates to $N^{2} - 1$ as required.

\section{Bosonic colour fields}
\label{SecTheoryB}
In the previous sections the anti-commuting nature of the additional colour fields meant that their Hilbert space was spanned by states transforming in representations built out of totally anti-symmetric tensor products. In turns out, as has been shown in \cite{Bastwl2}, that \textit{bosonic} colour fields can also provide the correct coupling to the gauge field. In the proceeding sections we will show that this choice of commuting fields instead supplies a basis of fully \textit{symmetric} representations of $SU(N)$ which will form the building blocks of the complete colour information associated to the matter field. We will be able to arrive at a similar formula to (\ref{main2}) based on this alternative approach to generating the Wilson loop interaction, which may be useful in instances where it is more natural -- or less awkward -- to use bosonic variables. Before we introduce this new approach it should be pointed out that it is already possible to generate fully symmetric representations using our previous results. Following the notation above, this can be done by using $F = p$ families of anti-commuting colour fields and projecting onto the sector of the Hilbert space with occupation numbers given by $n = (1, 1, \ldots, 1)$. The total number of degrees of freedom is easily calculated using the main formula (\ref{main2}) and turns out to be $\nCr{N\!+\!p\!-\!1}{p}$, coinciding with the dimension of the representation with $p$ fully symmetric indices. 

We prefer, however, to develop here a formalism where the fully symmetric representations are the fundamental constituents. To this end we define $N$ bosonic fields $\bar{c}^{r}$ and $c_{r}$ which transform in the (anti-)fundamental representation of $SU(N)$. These fields have the following Poisson brackets
\begin{equation}
	\{\bar{c}^{r}, c_{s}\}_{_{PB}} = i\delta^{r}_{s}\,; \qquad \{\bar{c}^{r}, \bar{c}^{s}\}_{_{PB}} = 0 = \{c_{r}, c_{s}\}_{_{PB}}~,
	\label{comm}
\end{equation}
that can be seen to follow from the particle action $S[\bar c,c]=\int_0^1 d\tau \bar c^r \dot c_r$,
and we use them to incorporate the colour indices of the gauge group generators since, defining 
\begin{equation}
	S^{a} = \bar{c}^{r}(T^{a})_{r}{}^{s}c_{s}~, 
	\label{als}
\end{equation}
one gets
\begin{align}
\{S^{a}, S^{b}\}_{_{PB}} = f^{abc}S^{c}~.
\end{align}
 We can therefore also use these commuting variables to represent the coupling between the particle and the gauge field and so take care of the non-commuting nature of the gauge group generators in a fully classical worldline action. This approach has been used successfully in worldline applications in the past where the same first order action  was used to enforce the path ordering inside functional integrals \cite{Bastwl2}. As before, this action is appended to the phase-space action (\ref{N=1}) along with the associated change to the conjugate momentum: $\pi^{\mu}$ is replaced by $\widetilde{\pi}^{\mu} = p^{\mu} - A^{a\mu} \bar{c}^{r}(T^{a})_{r}{}^{s}c_{s}$, which modifies the super-charge and Hamiltonian to 
\begin{equation}
\widetilde{Q} = \psi \cdot \widetilde{\pi}; \qquad	\widetilde{H} = \widetilde{\pi}^{2} + \frac{i}{2} \psi^{\mu}F^{a}_{\mu\nu}\psi^{\nu} \bar{c}^{r}(T^{a})_{r}{}^{s} c_{s}
\end{equation}
with $F_{\mu\nu}^{a}$ the components of the full (non-Abelian) field strength tensor with respect to the gauge group generators. 
To briefly examine the Hilbert space associated to the additional fields we promote $\bar{c}$ and $c$ to creation and annihilation operators $\hat{c}^{\dagger}$ and $\hat{c}$ and consider their action on \textit{bosonic} coherent states 
\begin{equation}
	\bstate{u} =\bvac e^{\bar{u}^{r} \hat{c}_{r}}; \qquad \bstate{u}\hat{c}^{\dagger r} = \bar{u}^{r}\bstate{u}; \qquad \bstate{u}\hat{c}_{r} = \partial_{\bar{u}^{r}}\bstate{u},
\end{equation}
where the $\bar{u}^{r}$ are the complex (commuting) eigenvalues of the $\hat{c}^{ \dagger r}$. The commutation relations are solved in this basis by taking $\hat{c}^{\dagger r} \sim \bar{u}^{r}$ and $\hat{c}_{r} \sim \partial_{\bar{u}^{r}}$. The Fock space is built by acting with creation operators on the vacuum and (in contrast to when the additional fields are Grassmann valued) is infinite dimensional. Indeed, wavefunctions $\Phi(x, \bar{u})$ now have a non-terminating expansion
\begin{equation}
	\Phi(x, \bar{u}) = \phi(x) + \phi_{r_{1}}(x) \bar{u}^{r_{1}} + \phi_{r_{1} r_{2}} \bar{u}^{r_{1}}\bar{u}^{r_{2}}(x) + \ldots + \phi_{r_{1}r_{2}\ldots r_{p}} \bar{u}^{r_{1}}\bar{u}^{r_{2}}\!\!\!...\bar{u}^{r_{p}} +\ldots
	\label{Focks}
\end{equation}
where the components transform in representations constructed out of $p$ fully \textit{symmetric} tensor products. The wavefunction thus transforms as a reducible sum. We resolve this as before, constructing a projector by gauging the $U(1)$ symmetry present in the worldline theory,  $c_r\to e^{-i\vartheta} c_r $, $\bar c^r\to \bar c^r e^{i\vartheta}$. This leads to a new action
\begin{equation}
	S\left[\omega, p, \psi, e, \chi, \bar{c}, c, a \right] = \int_{0}^{1} d\tau \, \bigg[ p \cdot \dot{\omega} + \frac{i}{2}\psi \cdot \dot{\psi} + i\bar{c}^{r} \dot{c}_{r} - e\widetilde{H} - i \chi \widetilde{Q}- a(L - s)\bigg],
	\label{Scas}
\end{equation}
where $a(\tau)$ is the gauge field which behaves under the $U(1)$ symmetry as $\delta a = \dot{\vartheta}$ and $L = \bar{c}^{r} c_{r}$ is the number operator for the colour fields which generates their $U(1)$ transformations. The Chern-Simons term, $S_{CS} = \int d\tau \, a(\tau) s$, is familiar by now and involves the constant, $s = n + \frac{N}{2}$. The equation of motion for $a(\tau)$ imposes the constraint $L = s$ which acts on the wavefunctions as
\begin{equation}
	\left(\hat{L} - \frac{N}{2}\right)\state{\Phi} = n\state{\Phi} \rightarrow \left(\bar{u}^{r}\frac{\partial}{\partial\bar{u}^{r}} - n \right)\Phi(x, \bar{u}) = 0,
\end{equation}
where this time we resolve the operator ordering ambiguity via symmetrisation so that $\hat{L} = \frac{1}{2}\left(\bar{u}^{r}\partial_{\bar{u}^{r}} + \partial_{\bar{u}^{r}}\bar{u}^{r}\right)$. With $\Phi$ given by (\ref{Focks}), this constraint picks out the component that transforms with exactly $n$ fully symmetrised indices, $\phi_{r_{1} r_{2} \ldots r_{n}}$. This brief analysis can be compared to the discussion in Section \ref{SecTheory}, where many similarities can be found.

\section{The generalised worldline theory}
\label{SecGeneral}
To describe representations other than those corresponding to fully symmetric tensor products requires us to generalise these ideas to include multiple families of the additional commuting fields. Following Section \ref{SecExtension} an additional index is appended to the colour fields to produce $F$ copies of these colour degrees of freedom, $\bar{c}^{r}_{f}$ and $c_{fr}$. The Poisson brackets are generalised to
\begin{equation}
	\{\bar{c}^{r}_{f}, c_{gs}\}_{_{PB}} = i\delta^{r}_{s}\delta_{fg}; \qquad \{\bar{c}^{r}_{f}, \bar{c}^{s}_{g}\}_{_{PB}}  = 0 = \{c_{fr}, c_{gs}\}_{_{PB}} 
	\label{commf}
\end{equation}
and a sum over these families of fields is included in each of the functions in the action (for example, $\widetilde{\pi}^{\mu} \rightarrow p^{\mu} - A^{a\mu} \bar{c}^{r}_{f}(T^{a})_{r}{}^{s}c_{fs}$). This again enlarges the $U(1)$ symmetry group to the $U(F)$ that rotates between the new fields. The $U(F)$ rotations are given by $c_{fr} \rightarrow \Lambda_{fg} c_{gr}$ and $\bar{c}^{r}_{f} \rightarrow \bar{c}^{r}_{g} \Lambda^\dagger_{gf}$ where $\Lambda :=e^{-i\lambda}$ and $\lambda$ is a constant Lie-algebra valued generator. We again introduce the generalised occupation numbers which are the conserved currents of this global symmetry, now built out of the bosonic colour fields as $L_{fg} = \bar{c}^{r}_{f}c_{fg}$. At the classical level, these currents obey the $U(F)$ algebra
\begin{equation}
	\left\{L_{ff^{\prime}}, L_{gg^{\prime}}\right\}_{_{PB}} = -i\delta_{f^{\prime}g}L_{fg^{\prime}} + i\delta_{fg^{\prime}}L_{gf^{\prime}},
\end{equation}
so they generate infinitesimal $U(F)$ transformations through Poisson brackets with $G = \lambda_{fg}L_{fg}$:
\begin{equation}
	\delta_{\lambda}c_{fr} = \left\{c_{fr}, G\right\}_{_{PB}} =- i\lambda_{fg}c_{gr}; \qquad \delta_{\lambda}\bar{c}^{r}_{f} = \left\{\bar{c}^{r}_{f}, G\right\}_{_{PB}} = i\bar{c}^{r}_{g}\lambda_{gf}.
	\label{deltaUfs}
\end{equation}

The Hilbert space is now substantially enlarged, consisting of components that transform in tensor products of the representation associated to each family and to achieve irreducibility it will be necessary to partially gauge this new symmetry. The richer Fock space can be expressed as a non-terminating sum over components of the form
\begin{equation}
	\Phi(x, \bar{u}) = \sum_{\{n_{1}, n_{2}, \ldots n_{\!F}\}} \phi_{r_{1} \ldots r_{n_{\!1}},s_{1} \ldots s_{n_{\!2}},\ldots, t_{1} \ldots t_{n_{\!F}}}(x)\bar{u}_{\!F}^{t_{1}}\!...\bar{u}_{F}^{t_{n_{\!F}}}\!\!\!...\bar{u}_{2}^{s_{1}}\!\!...\bar{u}_{2}^{s_{n_{\!2}}}\!\!\!\!...\bar{u}_{1}^{r_{1}}\!\!...\bar{u}_{1}^{r_{n_{\!1}}},
	\label{Psifs}
\end{equation}
where the indices are arranged into blocks of length $n_{f}$ which are fully symmetric. The form of the wavefunctions in (\ref{Psifs}) is nicely explained diagrammatically by using Young Tableaux notation:
\begin{equation}
\Yvcentermath1
	\Phi(x, \bar{u}) \sim \sum_{\{n_{1}, n_{2}, \ldots n_{\!F}\}}  \underbrace{{ \,\Small \yng(2) }\!\rhdots\!{ \Small\yng(2) }}_{n_{F}} \otimes  \ldots \otimes \underbrace{{ \Small \yng(2) }\!\rhdots\!{ \Small\yng(2) }}_{n_{2}}   \otimes \underbrace{{ \Small \yng(2) }\!\rhdots\!{ \Small\yng(2)}}_{n_{1}},
	\label{PsiTabs}
	\Yvcentermath0
\end{equation}
so that in general the wavefunction does not transform in an irreducible representation.

In order to have irreducibility, following the success of earlier sections, we gauge only a part of this $U(F)$ symmetry by introducing gauge fields $a_{fg}$ for $1 \leqslant g \leqslant f \leqslant F$ corresponding to a subset of the $L_{fg}$. We showed above that this subset satisfies a first class subalgebra and that it led to the desired projection onto a specified irreducible representation. To this end we modify the action to
\begin{align}
	S^{\prime}\left[\omega, p, \psi, e, \chi, \bar{c}, c, a\right] = \int_{0}^{1}\! d\tau \bigg[ p \cdot \dot{\omega} &+ \frac{i}{2}\psi \cdot \dot{\psi} + i\bar{c}^{r}_{f} \dot{c}_{rf} - e\widetilde{H} - i \chi \widetilde{Q}  \nonumber \\
	 &-\textstyle{\sum\limits_{f=1}^{F}}a_{f}\left(L_{f} - s_{f}\right) - \textstyle{\sum\limits_{g < f}}a_{fg}L_{fg} \bigg],
	\label{Smains}
\end{align}
which will form the basis for the remainder of this article. In equation (\ref{Smains}) we have split the $U(F)$ gauge fields into two types. As before, the diagonal generators $L_{f} \equiv L_{ff}$ are gauged by $a_{f} \equiv a_{ff}$ from which the off-diagonal elements have been separated and gauged by $a_{fg}$. To the former we have also introduced an independent Chern-Simons term associated to each family whose level is related to the desired occupation number of that sector of the Hilbert space. Invariance under local $U(F)$ transformations fixes the variation of the gauge fields to be
\begin{align}
	\delta_{\lambda} a_{fg}(\tau) &= \dot{\lambda}_{fg}(\tau) + i\sum_{g \leqslant k \leqslant f}\left( a_{fk}(\tau)\lambda_{kg}(\tau)-\lambda_{fk}(\tau)a_{kg}(\tau)\right)
	\label{deltaags}	
\end{align}
as is already familiar. 

We very briefly review how this achieves irreducibility, as the argument is essentially the same as in the case that the colour fields are anti-commuting. It suffices to consider the equations of motion implied by the gauge fields. These impose constraints on the physical state space
\begin{equation}
	\left(\hat{L}_{f} - \frac{N}{2}\right)\state{\Phi} = n_{f}\state{\Phi} \longrightarrow \left(\bar{u}^{r}_{f}\frac{\partial}{\partial \bar{u}_{f}^{r}} - n_{f}\right)\Phi(x, \bar{u}) = 0,
\end{equation}
where this time we have resolved an operator ordering ambiguity by symmetrising. This picks out from (\ref{Psifs}) the wavefunction component with $n_{f}$ symmetric indices associated to each family, $\phi_{r_{1} \ldots r_{n_{\!1}},s_{1} \ldots s_{n_{\!2}},\ldots, t_{1} \ldots t_{n_{\!F}}}(x)$. The remaining off-diagonal constraints then select only one irreducible representation from the tensor product denoted in (\ref{PsiTabs}). These constraints act on the Fock space as
\begin{equation}
	\hat{L}_{fg} \state{\Phi} = 0 ~\longrightarrow ~\bar{u}^{r}_{f}\frac{\partial}{\partial \bar{u}^{r}_{g}} \Phi(x, \bar{u}) = 0.
	\label{Lfgs}
\end{equation}
To illustrate the effect of this requirement we consider the same illustrative example as before, namely when $F = 2$ and $n_{2} = 2$, $n_{1} = 1$. In this case the constraints implied by $L_{1}$ and $L_{2}$ pick out the wavefunction component $\phi_{r_{1}r_{2},s_{1}}(x)$ so that the surviving part of $\Phi(x, \bar{u})$ can be written as a reducible sum
\begin{align}
	\Phi(x, \bar{u}) &= \phi_{r_{1}r_{2},s_{1}}(x) \bar{u}_{2}^{r_{1}}\bar{u}_{2}^{r_{2}}\bar{u}_{1}^{s_{1}} \nonumber \\
	&= \left[\phi_{(r_{1}r_{2},s_{1})}(x) + \frac{1}{3}\left(2\phi_{r_{1}r_{2},s_{1}}(x) - \phi_{s_{1}r_{1},r_{2}}(x) - \phi_{s_{1}r_{2},r_{1}}(x)\right) \right]\bar{u}_{2}^{r_{1}}\bar{u}_{2}^{r_{2}}\bar{u}_{1}^{s_{1}}
	\label{Psidecs}
\end{align}
which reflects the tensor product decomposition $\Yvcentermath1 {\Tiny \yng(2)} \otimes {\Tiny \yng(1)} = {\Tiny \yng(3)} \oplus {\Tiny \yng(2,1)} \Yvcentermath0$ (we have made use of the symmetries of the wavefunction to carry out some relabelling of indices to arrive at this). The surviving components must lie in the kernel of the $L_{fg}$, which act to remove a $\bar{u}^{r}_{g}$, to introduce a $\bar{u}^{r}_{f}$ and to symmetrise between indices. Indeed, acting with $\hat{L}_{21}$ on the creation operators above yields
\begin{align}
	\hat{L}_{21}\,\bar{u}^{r_{1}}_{2}\bar{u}^{r_{2}}_{2}\bar{u}^{s_{1}}_{1} &= \bar{u}_{2}^{t}\frac{\partial}{\partial \bar{u}^{t}_{1}}\,\bar{u}^{r_{1}}_{2}\bar{u}^{r_{2}}_{2}\bar{u}^{s_{1}}_{1} \nonumber \\
	&=\bar{u}^{s_{1}}_{2} \bar{u}^{r_{1}}_{2}\bar{u}^{r_{2}}_{2}
\end{align}
whose result is symmetric under interchange of the $r_{1}$, $r_{2}$ and $s_{1}$. Now the component in rounded brackets transforms under the representation with Young Tableau $\Yvcentermath1 { \Tiny \yng(2,1)} \Yvcentermath0$ which implies an antisymmetrisation between the $r_{1}$, $s_{1}$ and $r_{2}$, $s_{2}$ indices. For this reason, the only term that survives the action of $\hat{L}_{21}$ is $\phi_{(r_{1}r_{2},s_{1})}$ which the constraint therefore forces to vanish and the only surviving component is that which transforms in the representation with $n_{2}$ columns in the first row and $n_{1}$ columns in the second. 

The general case is similar to this example. The $L_{f}$ constraints pick out the wavefunction component with $n_{f}$ fully symmetrised indices associated to each family, which further breaks down into components which transform in the tensor product decomposition illustrated in the summand of (\ref{PsiTabs}). The $L_{fg}$ constraints further restrict attention to a single component which lies in their kernel. Given the form of the $L_{fg}$ this can only be achieved with a maximum amount of anti-symmetrisation between the indices of different families. Then the Young Tableau denoting the representation of the surviving component is 
\begin{equation}
\Yvcentermath1
	\Phi(x, \bar{u}) \sim  \left. \overset{\underset{\rvdots}{n_{f}}}{\underset{\overset{\rvdots}{n_{1}}}{\vphantom{\underset{ \Small \yng(4) }{ \overset{ \Small \yng(4) }{ \rvdots } }}}}\underset{ \Small \yng(4,2,1) }{ \overset{ \Small \yng(4,4,4) }{ \rvdots } } \overset{\overset{...}{\vphantom{ { \yng(1)} }}}{\underset{ \underset{...}{ \vphantom{  {\yng(1)}  } } }{...}} \vphantom{\underset{ \Small \yng(4,2,1) }{ \overset{ \Small \yng(4,4,4) }{ \rvdots } }} \underset{ \Small \yng(2) \hphantom{\yng(2)}  }{ \overset{ \Small \yng(4,3,2) }{ \rvdots } }\right\rbrace{ \Small \textrm{$F$ rows}}
\Yvcentermath0
\label{Psiirreps}
\end{equation}
which contains $n_{f}$ columns in the $f^{th}$ row with $n_{F} \geqslant n_{F-1} \geqslant \ldots \geqslant n_{2} \geqslant n_{1}$. Due to the anti-symmetrisation between the rows of each column implied by the shape of (\ref{Psiirreps}), only this component lies in the kernel of all of the $L_{fg}$. In this way, the partial gauging of the $U(F)$ symmetry provides a projection onto a single irreducible representation of $SU(N)$.  This complements our earlier work where the Tableau was made up by instead specifying the number of \textit{rows} in each of $F$ \textit{columns} and provides another approach which projects onto an arbitrarily chosen representation without being limited to those with single-row Tableaux. 

In the following section we will carry out the path integral quantisation of this theory for a particle whose path is closed. As discussed above this is relevant for calculations in the worldline approach to quantum field theory where it is related to the partition function of the quantum field theory describing a matter field interacting with a gauge field. We will calculate the degrees of freedom associated to the colour space of the matter field and compare the result to the dimension of the representation in (\ref{Psiirreps}).

\section{Path integral quantisation}
\label{SecPathB}
The situation is the same as in Section \ref{SecPath}: we consider a point particle that traverses a closed path by identifying its coordinates, $\omega^{\mu}$, at either end of the domain which therefore becomes $S^{1}$ as before. We proceed by integrating the momenta $p^{\mu}$ out so as to arrive (rotating to Euclidean signature) at a configuration space action
\begin{align}
	S\left[\omega, \psi, e, \chi, \bar{c}, c, a\right] =  \int_{0}^{1} d\tau \bigg[e^{-1}\dot{\omega}^{2}/2 + \frac{1}{2}\psi \cdot \dot{\psi} - \frac{\chi}{e} \dot{\omega} \cdot \psi + \bar{c}^{r}_{f} \dot{c}_{fr} &-i \bar{c}^{r}_{f} \mathcal{A}^{a}(T^{a})_{r}{}^{s} c_{fs}  \nonumber \\
	+ \textstyle{\sum\limits_{f=1}^{F}}ia_{f}\left(\bar{c}^{r}_{f}c_{fr}\! - \!s_{f}\right) &+ \textstyle{\sum\limits_{g < f}}ia_{fg}\bar{c}^{r}_{f}c_{gr} \bigg],
	\label{Smaincs}
\end{align}
where $s_{f}$ are the Chern-Simons level and we recall the previously defined worldline function $\mathcal{A} = \dot{\omega} \cdot A - \frac{e}{2}\psi^{\mu}F_{\mu\nu}\psi^{\nu}$ which parameterises the phase of the Wilson loop, $W = \mathscr{P} e^{i\int_{0}^{1} \mathcal{A}(\tau) d\tau}\,$. The object in question is the functional integral over configurations 
\begin{equation}
	\int \frac{\mathscr{D}e \mathscr{D} \chi \mathscr{D}\omega \mathscr{D}\psi \mathscr{D}\bar{c}\mathscr{D}c\mathscr{D}a}{\rm Vol(Gauge)} \, \exp{\left(-S\left[\omega, \psi, e, \chi, \bar{c}, c, a\right]\right)}.
	\label{Paths}
\end{equation}
The introduction of the additional bosonic fields now replaces the path ordering prescription. We take periodic boundary conditions on $\omega$, $e$ and $a$, but now also on on $\bar{c}$, $c$. The remaining fields, $\psi$ and $\chi$, have anti-periodic boundary conditions as usual. We follow the analysis employed in Section \ref{SecPath} by restricting attention to the colour degrees of freedom, considering the reduced path integral
\begin{align}
	\frac{1}{\rm Vol(Gauge)}\int & \prod_{f,r}\mathscr{D}\bar{c}_{f}^{r}\mathscr{D}c_{fr} \prod_{h < g}\mathscr{D}a_{gh} \prod_f\mathscr{D}a_f \nonumber \\&\times \exp{\left(-\int_{0}^{1} d\tau \, \left[  \textstyle{\sum\limits_{f=1}^{F}} \left(\bar{c}^{r}_{f} D_{f} c_{fr} - is_{f}a_{f}\right) +  \textstyle{\sum\limits_{h < g}}ia_{fg}\bar{c}^{r}_{f}c_{gr} \right]\right)},
	\label{pics}
\end{align}
where the diagonal gauge fields are absorbed into defining a covariant derivative $D_{f} \equiv \left(\frac{d}{d\tau} + ia_{f}\right)$ and we have dropped the coupling to the external gauge field. We will show that this functional integral evaluates to the correct number of degrees of freedom associated to the colour space for any choice of $s = (s_{1}, s_{2}, \ldots , s_{F})$. In the worldline formalism we would also have to include the functional integrals involving the embedding coordinates, $\omega$, the spin degree of freedom, $\psi$ and the worldline supergravity multiplet $e$ and $\chi$, which would allow us to calculate scattering amplitudes and other physical observables. For the present article we will continue to deal only with the colour degrees of freedom, leaving to future work the inclusion of the interaction with the gauge field.

The action of the colour fields in (\ref{pics}) is of course still invariant under the auxiliary symmetry group discussed above. As before, we gauge fix this symmetry by fixing the fields $a_{fg}$ to be the constant matrix 
\begin{equation}
	\hat{a}_{fg} = \begin{pmatrix}
		\theta_{1} & 0 & \cdot & 0 \\ 0 & \theta_{2} & \cdot & 0 \\ \cdot & \cdot & \cdot & \cdot \\ 0 & 0 & \cdot & \theta_{F}
	\end{pmatrix}.
	\label{afixs}
\end{equation}
where the parameters are interpreted as angles in $[0, 2\pi]$ and are the moduli to be integrated over. Indeed, integrating over these variables will provide the projection onto an irreducible representation of $SU(N)$. We have already found the Faddeev-Popov determinant associated to this gauge fixing which led to the measure on the the moduli, (\ref{mu}), which takes the form
\begin{equation}
	\mu\left(\{\theta_{k}\}\right) = \prod_{h < g}2i \sin{\left(\frac{\theta_{g} - \theta_{h}}{2}\right)}
\end{equation}
and is the correct factor ensuring gauge invariance of functional integrals on our chosen gauge slice. This measure is also the factor that is required to ensure a projection onto a single irreducible representation of the symmetry group. 

\subsection{Evaluation of the path integral}
The gauge fixed path integral takes the form
\begin{align}
	K_{F}\prod_{f = 1}^{F} \int _{0}^{2\pi} \frac{d\theta_{f}}{2\pi} \int \mathscr{D} (\bar{c},c)  \,\mu\left(\{\theta_{k}\}\right) \exp{\left(- \int_{0}^{1}\!d\tau \sum_{f = 1}^{F} \left( \bar{c}^{r}_{f} D_{f} c_{fr} - i\theta_{f}s_{f} \right) \right)}
\end{align}
where the covariant derivative takes the simple form $D_{f} = \left(\frac{d}{d\tau} + i\theta_{f}\right)$ and $K_{F}$ is the normalisation constant which also appeared for fermionic colour fields. The integration over $\bar{c}$ and $c$ leads to a functional determinant which is defined as the product of its non-zero eigenvalues:
\begin{equation}
	\prod_{f = 1}^{F}\left[\underset{\textrm{\Tiny \!\!PB}}{\Det}\left(\frac{d}{d\tau} + i\theta_{f}\right)\right]^{-N} =\, \prod_{f = 1}^{F}\left(2i \sin{\left(\frac{\theta_{f}}{2}\right)}\right)^{-N},
\end{equation}
where we have made use of the results of Appendix~\ref{sec:determinants}. This gives a formula for the number of degrees of freedom associated to the colour space, which is the main result of this section,
\begin{equation}
	K_{F}\prod_{f = 1}^{F} \int _{0}^{2\pi} \frac{d\theta_{f}}{2\pi}e^{i\theta_{f}s_{f}}\mu\left(\{\theta_{k}\}\right)\left(2i \sin{\left(\frac{\theta_{f}}{2}\right)}\right)^{-N}.
	\label{mains}
\end{equation}
We will now give some example calculations making use of this formula to demonstrate that it provides the expected results.

The simplest example to consider is for $F = 1$, whereby the formula is familiar from previous work and determines the dimension of fully symmetric representations. Taking $s_{1} = n_{1} + \frac{N}{2}$ and using $\mu(\theta_{1}) = 1$, $K_{1} = 1$, (\ref{mains}) becomes
\begin{align}
		&\int_{0}^{2\pi} \frac{d\theta_{1}}{2\pi} e^{i\theta_{1} n_{1}}e^{i\theta_{1}\frac{N}{2}}\left(e^{i \frac{\theta_{1}}{2}} - e^{-i \frac{\theta_{1}}{2}}\right)^{-N} \nonumber \\
	=&\int_{0}^{2\pi} \frac{d\theta_{1}}{2\pi} e^{i\theta_{1} n_{1}}\left(1 - e^{-i\theta_{1}}\right)^{N}.
\end{align}
It so far has proven convenient to make a change of variable to write our expressions in terms of the $U(1)$ Wilson-loop on the circle, $z_{1} = e^{i\theta_{1}}$. Doing so we find
\begin{align}
	&\oint \frac{dz_{1}}{2\pi i z_{1}} \frac{z_{1}^{n_{1}}}{(1 - \frac{1}{z_{1}})^{N}} \nonumber \\
	=&\oint \frac{dz_{1}}{2\pi i}\frac{z_{1}^{N-1 + n_{1}}}{\left(z_{1} - 1\right)^{N}}.
\end{align}
The contour of integration is the circle $\left|z_{1}\right| = 1$, deformed to enclose the pole on the real axis at $z_{1} = 1$. This follows from the regularisation procedure outlined in Appendix B, which requires a shift $\theta_{1} \rightarrow \theta_{1} - i\epsilon$, sending $z_{1} \rightarrow z_{1}e^{\epsilon}$ for a small positive parameter $\epsilon$. The integral is easily calculated to be
\begin{equation}
	\oint \frac{dz_{1}}{2\pi i}\frac{z_{1}^{N + n_{1}}}{\left(z_{1} - 1\right)^{N}} = \nCr{N\!+\!n_{1}\! -\!1}{n_{1}},
	\label{contours}
\end{equation}
which agrees with the dimension of the $SU(N)$ representation with $n_{1}$ fully symmetrised indices. This is in agreement with the results of \cite{Bastwl2}, where scattering amplitudes were calculated in the worldline formalism involving matter fields which transformed in fully symmetric representations of $SU(N)$. 

We also return to (\ref{Psidecs}) where the calculation is more interesting. This requires the use of $F = 2$ families of fields and we take the general case of occupation numbers $n = (n_{1}, n_{2})$ where $n_{1} \leqslant n_{2}$. Similarly to Section \ref{SecPath}, this is enforced by Chern-Simons terms $s_{1} = n_{1} + \frac{N}{2} - \frac{1}{2}$ and $s_{2} = n_{2} + \frac{N}{2} + \frac{1}{2}$, in which case (\ref{mains}) gives (recall that $K_{2} = 1$)
\begin{equation}
	\int_{0}^{2\pi}\frac{d\theta_{1}}{2\pi}\int_{0}^{2\pi}\frac{d\theta_{2}}{2\pi} e^{i \theta_{1}\left(n_{1}-\frac{1}{2}\right)}e^{i \theta_{2}\left(n_{2}+\frac{1}{2}\right)}\left(1 - e^{-i\theta_{1}}\right)^{-N}\left(1 - e^{-i\theta_{2}}\right)^{-N}\left(e^{i \frac{\theta_{1}}{2}}e^{-i\frac{\theta_{2}}{2}} - e^{-i\frac{ \theta_{1}}{2}}e^{i\frac{\theta_{2}}{2}}\right).
\end{equation}
In this expression we have extracted factors of $e^{-i\theta_{1}\frac{N}{2}}$ and $e^{-i\theta_{2}\frac{N}{2}}$ from the first two terms in rounded brackets to cancel them off against their counterparts for simplification. We will again make use of the Wilson-loop variables by defining $z_{1} = e^{i\theta_{1}}$ and $z_{2} = e^{i\theta_{2}}$ which allows us to re-express this formula as a multiple integral on the complex plane:
\begin{equation}
	\oint\!\oint \frac{d z_{1}}{2\pi i}\frac{dz_{2}}{2\pi i}\left(\frac{z_{1}^{N-1 + n_{1}}}{(z_{1} - 1)^{N}}\frac{z_{2}^{N-1 + n_{2}}}{(z_{2} -1)^{N}} -\frac{z_{1}^{N-1 + (n_{1} - 1)}}{(z_{1} - 1)^{N}}\frac{z_{2}^{N - 1 + (n_{2} + 1)}}{(z_{2} - 1)^{N}}\right).
\end{equation}
Making use of (\ref{contours}) this evaluates to
\begin{align}
	&\nCr{N \!+\! n_{1}\! -\! 1}{n_{1}}\nCr{N\! + \! n_{2} \! - \! 1}{n_{2}} - \nCr{N \!+\! n_{1}\! -\! 2}{n_{1} \!-\! 1} \nCr{N \!+ \!n_{2}}{n_{2} \! + \! 1} \nonumber \\
	&= \, \frac{(N + n_{2} - 1)!}{n_{2}!(N-1)!}\frac{(N + n_{1} - 2)!}{(N - 2)!n_{1}!}\frac{n_{2} + 1 - n_{1}}{n_{2} + 1} \nonumber \\
	&=\, \nCr{N \! + \! n_{2} \! -\!1}{n_{2}}\nCr{N \! + \! n_{1} \!-\! 2}{n_{1}}\frac{n_{2} + 1 - n_{1}}{n_{2} + 1}.
	\label{dof2s}
\end{align}
It is simple to verify that this corresponds to the dimension of the representation of $SU(N)$ with Young Tableau
\begin{equation}
	\Yvcentermath1
	^{n_{2}}_{n_{1}} {\Small \yng(3,3)} \!\rhdots\! {\Small \yng(3,1)}
	\Yvcentermath0
\end{equation}
having $n_{2}$ columns in the first row and $n_{1}$ columns in the second\footnote{Following \cite{Georgi} this Tableaux has ``factors'' $\frac{(N + n_{2} - 1)!}{(N - 1)!}\frac{(N + n_{1} - 2)!}{(N - 2)!}$ and ``hooks'' $\frac{n_{1}!(n_{2} + 1)!}{(n_{2} + 1 - n_{1})!}$ whose quotient gives the dimension in (\ref{dof2s}).}. Specifying $n_{2} = 2$ and $n_{1} = 1$ projects on the representation $\Yvcentermath1 {\Tiny \yng(2,1)} \Yvcentermath0$ as in (\ref{Psidecs}). 

As before it is possible to write our main formula more compactly by representing all of the $U(F)$ moduli by Wilson-loop variables, defining $z_{f} = e^{i\theta_{f}}$ for $f \in [1, \ldots F]$. Assuming occupation numbers $n = \left(n_{1}, n_{2}, \ldots, n_{F}\right)$ we take Chern-Simons levels $s_{f} = n_{f} + \frac{N+1}{2} - f$. Then following similar lines to before the number of degrees of freedom carried by the matter field in the worldline approach can be cast into the form
\begin{equation}
	\prod_{f = 1}^{F}\oint \frac{dz_{f}}{2\pi i} \prod_{h < g}\left(1 - \frac{z_{h}}{z_{g}}\right)\frac{z_{f}^{N - 1 + n_{f}}}{(1 - z_{f})^{N}} \,,
	\label{main2s}
\end{equation}
where each integration contour encloses the pole at $z = 1$. We have verified that this produces the correct number of degrees of freedom for any representation of $SU(N)$ whose Young Tableau has $n_{f}$ columns in the $f^{th}$ row. The formula (\ref{main2s}) is used by fixing the numbers in $n$ and expanding the product into a sum of multiple complex integrals. For example, with $F = 6$ and $n = (1, 2, 2, 4, 6, 7)$ (assuming that $N \geqslant 6$), the integration provides
\begin{equation}
\Yvcentermath1
	\dim \, {\Small \yng(7,6,4,2,2,1)}.
	\Yvcentermath0
\end{equation}
This shows that the modular measure $\mu(\{\theta_{k}\})$ correctly projects onto the wavefunction component transforming in the desired irreducible representation of the gauge group. This measure follows from the partial gauging of the $U(F)$ symmetry that existed in the worldline theory. In the worldline approach it would be necessary to also complete the remaining integrals over the matter and supergravity fields in (\ref{Smaincs}). This procedure is well known and gives the dynamical and spin degrees of freedom of the particle. In future work we intend to also include the coupling to the gauge field in order to show that the additional colour fields produce the expected interaction -- this would be the Wilson-loop coupling for the chosen representation of $SU(N)$. 

As we have already mentioned one of the most physically significant representations that one could project onto is the adjoint of $SU(N)$. It remains possible to achieve this in the current setting: we would need $F = N - 1$ families of fields and should take the vector $n = (2, 1, \ldots 1)$. We have verified that formula (\ref{main2s}) then evaluates to $N^{2} - 1$ as desired. Finally, before introducing the bosonic colour fields we commented that fully symmetric representations can be constructed out of a theory based on fully anti-symmetric representations. The converse is also true; choosing $0 \leqslant F = p\leqslant N$ and $n = (1, 1, \ldots 1)$, our main formula gives the correct number of degrees of freedom of an $SU(N)$ tensor with $p$ fully antisymmetric indices, $\nCr{N}{p}$, and vanishes if $F > N$.

\section{Conclusion}
We have presented the construction of two related worldline theories which describe a spin 1/2 matter field transforming in an arbitrary representations of $SU(N)$. We have shown how coupling the matter field to families of either commuting or anti-commuting auxiliary fields replaces the awkward path ordering prescription and allows the description of richer matter content, including tensors with mixed symmetry. We examined the Hilbert space of the auxiliary worldline fields which carry the colour information of the matter field and described how to isolate wavefunction components transforming in a desired irreducible representation. In the case that the auxiliary fields are anti-commuting, they provide a Fock space whose states transform in tensor products of  $F$ anti-symmetric representations of $SU(N)$, whereas their commuting counterparts span a Hilbert space with states that are tensor products of $F$ symmetric representations. The projection on an irreducible representation was achieved by partially gauging a $U(F)$ symmetry which rotates between the $F$ families of additional fields, and the introduction of Chern-Simons terms to constrain the occupation numbers of the particle wavefunction. Our main result arose from path integral quantisation of the worldline theory which provided a compact formula for the colour degrees of freedom of the matter field. 

In future work we will include the coupling to the gauge field in order to complete a worldline description of arbitrary matter multiplets coupled to an external non-Abelian field. This will be done by reinstating the Wilson-loop coupling between the colour fields and the gauge fields and computing the resulting path integral: this will provide a complete framework for the computation of gluon amplitudes in the presence of any form of matter (note that chiral fermions can also be described in a first quantised setting by following the ideas presented in \cite{Me1, Paul2}). The utility of this approach is clear from the existing efficiency of the worldline formalism, which ought to be preserved by the introduction of the colour fields. In general, the worldline formalism leads to Bern-Kosower rules that simplify the calculation of scattering amplitudes. In the non-Abelian case, these rules state that in order to achieve the full gauge invariance of the on-shell gluon amplitudes, one must append the one-particle irreducible expressions, that naturally come from the worldline representation, with ``tree replacement rules'' that allow the inclusion of gluon self-interactions.  We wish to use our formalism to generalise Bern-Kosower rules for gluon scattering in the presence of an arbitrary matter field. 

Although we have limited our attention to closed worldlines for the sake of simplicity, these methods are also well-suited for application to tree-level amplitudes, where the particle endpoints are fixed~\cite{Me2un, Ahmadiniaz:2015xoa}. This has application to dressed propagators in quantum field theory and the analysis of bound states~\citep{Bastianelli:2014bfa}. Finally, the old link between the worldline approach and string theory has recently been re-examined in \cite{Us1, Us2}, where a contact interaction was introduced on the worldsheets of tensionless spinning strings. Identifying the endpoints of the strings with particle worldlines reproduced the Wilson-loop interaction for spinor QED. Generalising this theory to non-Abelian symmetry groups would require the incorporation of the colour degrees of freedom onto the string worldsheet and would lead to an alternative description of particle interactions in terms of interacting coloured strings. Furthermore, it would be worthwhile examining the relative merits of using the bosonic colour carrying fields compared to the Grassmann auxiliary fields as it would be useful to know which approach tends to offer the most simplicity and versatility for the purpose of calculations in the worldline approach. 

Much work has also been carried out using worldline techniques for fields that are coupled to the gravitational field. The techniques we have presented here can be incorporated into that context with minor modifications and may provide a powerful approach to the calculation of amplitudes involving gluons and/or gravitons. 

\subsection*{Acknowledgements}
The authors would like to express their deepest thanks to the warmth and hospitality of INFN Bologna where this work began life. The progress of this research has benefited greatly from the insight of Fiorenzo Bastianelli and Christian Schubert with whom the authors have enjoyed many helpful discussions which have had direct bearing on the material presented in this article. JPE would thank INFN Bologna and University of Modena and Reggio Emilia for additional financial support to make this manuscript possible. 

\appendix
\section{$U(F)$ Gauge Fixing} 
\label{sec:appA}
Here we present the fixing of the non-Abelian $U(F)$ symmetry discussed in the main text. The generator of the symmetry, $\lambda$, is in the Lie algebra of $U(F)$ and produces the following finite transformation on the gauge field:
\begin{equation}
	a \rightarrow U^{-1}aU -iU^{-1}\dot{U}; \qquad U(\tau) = e^{i\lambda(\tau)}.
\end{equation}
It is tempting to try to use such a transforming to fix the gauge field to vanish identically. Simple algebra shows this is equivalent to solving $\frac{d}{d\tau}U(\tau) = -ia(\tau)U(\tau)$ which can be done by making use of the Path ordering prescription
\begin{equation}
	U(\tau) = \mathscr{P}e^{-i \int_{0}^{\tau} a(\tau')d\tau' }.
\end{equation}
Since the domain is $S^{1}$ we require the transformation to be periodic, so this solution is not admissible. Instead, we can define a constant group valued matrix $e^{-i\Theta} = \mathscr{P}e^{-i \int_{0}^{1} a(\tau)d\tau}$ and construct a gauge transformation which takes $a_{fg}(\tau)$ onto the constant $\Theta$ in the Lie algebra of the symmetry group: 
\begin{equation}
	\tilde{U}(\tau) = U(\tau)e^{i\Theta} \Rightarrow a_{kj}(\tau) \rightarrow \Theta_{kj}.
\end{equation}
With the partial gauging we use in this paper, the $\Theta$ will be upper-triangular. However, when we return to the functional integral over the colour fields, the resulting functional determinant is easily shown to be independent of the off diagonal terms. These then factor out of the path integral, leading to an overall constant which is cancelled upon normalisation. So it suffices to keep only the diagonal elements of the Cartan subalgebra which leaves $\hat{a}_{kj}$ in (\ref{afix}). These remaining moduli can be identified as angles by considering large $U(1)^{F} \subset U(F)$ transformations and requiring periodicity. 

The final ingredient is the integration measure on the moduli, which we derive using the Faddeev-Popov formalism. To avoid overcounting one factorises the integration over orbits of the gauge group out of the functional integration. This is achieved by integrating only over the moduli $\theta_{k}$, compensating for this gauge fixing by introducing the ghosts in (\ref{Sghost}) That is,
\begin{equation}
	\int \mathscr{D}a \rightarrow \int \mathscr{D}U \int \mathscr{D} a \int \mathscr{D}(\bar{\gamma}, \gamma) \, e^{-S_{g}\left[ \bar{\gamma}, \gamma, \{\theta\}\right]} \delta\left(a^{U} - \hat{a}\right)
\end{equation}
where $\hat{a}$ is the gauge fixed form of $a$ and $a^{U}$ represents the gauged transformed value of $a$. In the main text we used the results of Appendix B to integrate over the ghost degrees of freedom which, cancelling the functional integration over $U$ with the volume of the gauge group, provided gauge fixed integrals of the form
\begin{equation}
	\int \mathscr{D} a \, \Omega[a] \rightarrow \prod_{k = 1}^{F} \int \frac{d\theta_{k}}{2\pi} \mu(\{\theta_{k}\})\, \Omega[\hat{a}(\{\theta_{k}\})]
\end{equation}
where $\Omega$ is any functional of the gauge fields and $\mu$ denotes the Faddeev-Popov measure maintaining gauge invariance.

\section{Functional Determinants}
\label{sec:determinants}
In this appendix we calculate the various functional determinants which arose in the main text. The main difference between the situations that are encountered is in the boundary conditions imposed on the fields in the functional integration. We define the functional determinant
\begin{equation}
	\Det \left(\frac{d}{d\tau} + i\theta\right) 
\end{equation}
as the product of non-zero eigenvalues of the operator in brackets. On the space of periodic functions, the eigenvalue equation
\begin{equation}
	\left(\frac{d}{d\tau} + i\theta\right)f(\tau)  = \mu f(\tau)
\end{equation}
is solved by eigenfunctions $f(\tau) = e^{-i \theta \tau}e^{i \mu \tau}$, requiring $\mu = \theta + 2n\pi$ for $n \in \mathbb{Z}$. We arrange the product over these eigenvalues by pairing up the positive and negative integers (taken as part of our regularisation of the infinite product) as
\begin{equation}
	\theta \prod_{n =1 }^{\infty} \left(1 - \frac{\theta^{2}}{\left(2 n \pi\right)^{2}}\right) \prod_{n = 1}^{\infty} \left(2 n \pi\right)^{2}.
\end{equation}
The latter product can be $\zeta$-function regularised and evaluates to a constant, independent of $\theta$ (see \cite{Me2un}). The first product is well-known from the expansion of the $\sin$ function so we arrive at
\begin{equation}
	\underset{\textrm{\Tiny\!\!PB}}{\Det} \left(\frac{d}{d\tau} + \theta\right) = 2 i \sin{\left(\frac{\theta}{2}\right)}.
\end{equation}
For anti-periodic boundary conditions the eigenvalues are modified to $\mu = \theta + (2n + 1)\pi$, so their product becomes
\begin{equation}
	\theta \prod_{n =0 }^{\infty} \left(1 - \frac{\theta^{2}}{\left(\left(2 n +1\right)\pi\right)^{2}}\right) \prod_{n = 0}^{\infty} \left(\left(2 n+1\right) \pi\right)^{2}.
\end{equation}
The final product gives an overall constant and the first is related to the infinite product expansion of the $\cos$ function, so we find
\begin{equation}
	\underset{\textrm{\Tiny\!\!APB}}{\Det} \left(\frac{d}{d\tau} + \theta\right) = 2 \cos{\left(\frac{\theta}{2}\right)}.
\end{equation}
These results are used extensively in the main text.

We also wish to return to the calculation of the determinant on the space of periodic functions as the regularisation is important for the discussion presented in the main text. For this reason we re-derive the result using techniques derived from canonical quantisation. The determinant arising out of the following integration can be expressed as a trace over states in the Hilbert space:
\begin{align}
	\left(\underset{\textrm{\Tiny\!\!PB}}{\Det} \left(\frac{d}{d\tau} + \theta\right)\right)^{-1} &= \oint_{\textrm{PB}}\mathscr{D} (\bar{c},c)\, e^{-\int_{0}^{1} d\tau \, \bar{c} \left(\frac{d}{d\tau} + \theta\right) c } \nonumber \\
	&= \tr\, e^{- i\frac{\theta}{2}\left(\hat{c}^{\dagger} \hat{c} + \hat{c}\hat{c}^{\dagger}\right)}. 
\end{align}
The trace is most easily evaluated in the occupation number basis, where the states are eigenvalues of $\hat{n} = \hat{c}^{\dagger}\hat{c}$. This leads to
\begin{align}
	\left(\underset{\textrm{\Tiny\!\!PB}}{\Det} \left(\frac{d}{d\tau} + \theta\right)\right)^{-1} &= \sum_{n = 0}^{\infty}\left<n\right|e^{-i \frac{\theta}{2}\left(\hat{n} + \frac{1}{2}\right)}\state{n} \nonumber \\
	&= \sum_{n = 0}^{\infty}e^{-i \frac{\theta}{2}\left(n + \frac{1}{2}\right)}.
\end{align}
We are obliged to regularise the geometric series by taking $\theta \rightarrow \theta - i\epsilon$ and arrive at
\begin{equation}
	\left(\underset{\textrm{\Tiny\!\!PB}}{\Det} \left(\frac{d}{d\tau} + \theta\right)\right)^{-1} = \frac{e^{-i\frac{\theta}{2}}}{1 - e^{-i\theta}}  = \left(2 i \sin{\left(\frac{\theta}{2}\right)}\right)^{-1}.
\end{equation}
The necessary regularisation of $\theta$ affects the Wilson-loop variable $z = e^{i\theta} \rightarrow z = ze^{\epsilon}$, which is important in determining the contour in the complex plane along which this parameter is integrated. This changes the original circle of radius $\left|z\right| = 1$ to a circle that is slightly larger, therefore encompassing any poles that lie at unit length from the origin. On the other hand, when the variables $\bar{c}$ and $c$ are Grassmann valued then the trace becomes
\begin{equation}
	\underset{\textrm{\Tiny\!\!APB}}{\Det} \left(\frac{d}{d\tau} + \theta\right) = \sum_{n = 0, 1} e^{-i\theta\left(n - \frac{1}{2}\right)} = 2\cos{\left(\frac{\theta}{2}\right)}
\end{equation}
which does not require such regularisation.
\bibliographystyle{JHEP}
\bibliography{bibSym}
\end{document}